\documentclass[usenatbib]{mn2e}  

\pdfoutput=1 

\usepackage[pdftex]{graphicx} 
\usepackage{times}
\usepackage{rotate} 
\usepackage{amssymb} 
\usepackage{rotating}
\usepackage{epstopdf} 
\usepackage{graphicx}
\usepackage{array}

\title[X-ray AGN clustering with photometric redshifts]{Measuring the Dark Matter Halo Mass  of X-ray AGN at z$\sim$1 using photometric redshifts}

\author[Mountrichas et al.]{G. Mountrichas$^1$, A. Georgakakis$^{2,1}$,  A. Finoguenov$^{3,2}$, G. Erfanianfar$^2$, M.   C.    Cooper$^4$, 
\newauthor A.  L.  Coil$^5$, E. S. Laird$^6$, K.  Nandra$^2$, J. A. Newman$^7$\\ \\ 
$^1$National Observatory of Athens, V.  Paulou  \& I.  Metaxa, 11532,  Greece\\ 
$^2$Max Planck  Institut f\"{u}r Extraterrestrische  Physik, Giessenbachstra\ss e, 85748 Garching, Germany\\ 
$^3$Department of Physics, University of Helsinki, Gustaf Hallstromin katu 2a, FI-00014 Helsinki, Finlanda\\
$^4$Center for  Galaxy Evolution, Department  of  Physics and  Astronomy, University  of California,  Irvine, 4129
Frederick Reines Hall, Irvine, CA 92697, USA\\ 
$^5$Department of Physics and Center  for Astrophysics  and Space Sciences, University  of California, San
Diego,   9500  Gilman   Dr.,  La   Jolla,  CA   92093\\  
$^6$Astrophysics Group, Blackett Laboratory, Imperial College, Prince
Consort Rd , London SW7 2AZ, UK\\
$^7$University of Pittsburgh, Physics \& Astronomy Department, 3941 O'Hara Street Pittsburgh, PA, 15260, USA}

\begin{document}

\maketitle

\label{firstpage}

\begin{abstract}  Data from the  AEGIS, COSMOS  and ECDFS  surveys are
combined  to infer  the bias  and dark  matter halo  mass  of moderate
luminosity [$L_X (\rm 2-10\,keV)= 42.9 \, erg \, s^{-1}$] X-ray AGN at
$z\approx1$  via their cross-correlation  function with  galaxies.  In
contrast to standard cross-correlation function estimators, we present
a  method  that  requires  spectroscopy  only for  the  AGN  and  uses
photometric redshift  probability distribution functions  for galaxies
to  determine the  projected  real-space AGN/galaxy  cross-correlation
function.  The  estimated dark  matter halo mass  of X-ray AGN  in the
combined   AEGIS,   COSMOS   and   ECDFS   fields   is   $\approx   13
h^{-1}$\,M$_{\odot}$,  in agreement with  previous studies  at similar
redshift and  luminosity ranges.  Removing  from the sample the  5 per
cent of  the AGN  associated with X-ray  selected groups results  in a
reduction  by about  0.5\,dex in  the  inferred AGN  dark matter  halo
mass. The distribution  of  AGN in  dark  matter  halo mass  is
  therefore skewed  and the bulk  of the population lives  in moderate
  mass  haloes.  This  result favour  cold gas  accretion as  the main
  channel of supermassive black hole growth for most X-ray AGN.
\end{abstract}


\section{Introduction}

The  clustering  properties of  Active  Galactic  Nuclei  (AGN) are  a
powerful diagnostic  of the mechanism  that dominates the  fuelling of
Super-Massive  Black  Holes  (SMBHs)  across cosmic  time.   Cold  gas
accretion scenarios for example, predict  that AGN live in Dark Matter
Haloes (DMHs) of up to  few times $\rm 10^{12} \, h^{-1} \,M_{\odot}$,
almost  independent  of   accretion  luminosity  and  redshift  \citep
{Marulli2008, Hopkins2008a}.   In contrast, if the growth  of SMBHs is
dominated  by accretion  from a  hot quasi-hydrostatic  halo  then one
might  expect more luminous  AGN in  more massive  DMHs \citep[e.g.][]
{Fanidakis2012}.

Numerous observational programmes have  been initiated in the last few
years to  test those predictions by  measuring the DMH mass  of AGN as
function  of redshift and  accretion luminosity.   Large spectroscopic
campaigns of powerful UV-bright QSOs showed that this population lives
in DMHs  of a few times  $\rm 10^{12} h^{-1}\, M_{\odot}$  over a wide
range of redshifts \citep[e.g.][] {Croom2005, daAngela2008, Ross2009}.
This is consistent  with the predictions of cold  gas accretion models
for  the fuelling  of the  SMBHs.   At the  same time  there has  been
progress in  the study  of the clustering  of the less  luminous X-ray
selected  AGN. These systems  provide a  more representative  and less
biased  census  of  the  active  SMBH population  that  dominates  the
accretion  density of the  Universe \citep[e.g][]{Aird2010}.   The DMH
masses  of this  class of  sources are  estimated to  be,  on average,
larger  than   those  of  UV-bright   QSOs,  $\log  M_{DMH}=12.5-13.5$
\citep[$\rm  h^{-1}\,M_{\odot}$,][]{Cappelluti2012}.   This  has  been
interpreted as  evidence against cold-gas  accretion via major-mergers
in  those   systems  \cite*[e.g.][]  {Allevato2011,  Mountrichas2012}.
However, there is scatter among the individual measurements of the DMH
mass of  X-ray AGN,  which makes  it hard to  comment on  the redshift
and/or accretion luminosity dependence of their clustering properties.
This  is because  both random  errors, related  to e.g.   small sample
sizes, and  systematic uncertainties, such as  sample variance, affect
current determinations  of the  typical DMH mass  of X-ray  AGN.  Both
these  issues  are  related  to  the  strong  requirement  of  current
clustering  estimators  for  spectroscopy  to  get  robust  clustering
measurements.

The  real-space cross-correlation  function of  AGN with  galaxies for
example, is arguably  one of the most reliable  methods for clustering
studies   \cite*[e.g.][]   {Coil2009,   Mountrichas2012,   Krumpe2010,
Krumpe2012}.  Random errors are significantly suppressed when counting
AGN/galaxy pairs and the impact  of sample variance, which is shown to
affect  even   relatively  large-area  X-ray   surveys  \citep[e.g.][]
{Gilli2009}  is minimised.  However,  these features  do not  come for
free.    The   classic   method   for   determining   the   AGN/galaxy
cross-correlation  function requires  extensive spectroscopy  for both
AGN and galaxies.  This limits the number of X-ray fields that such an
analysis can be  applied, as spectroscopy for large  galaxy samples is
expensive  in resources.   One way  to address  this limitation  is to
relax the requirement for spectroscopy in clustering studies.

An important development  in the last few years  in this direction has
been the rapid improvement in  the quantity and quality of photometric
redshift   measurements,  particularly  for   galaxies  \citep[e.g.][]
{Coupon2009}. Large scale  multi-waveband photometric surveys, such as
the  SDSS (Sloan  Digital  Sky Survey,  \citealt {Abazajian2009})  and
CFHTLS  \citep  {Coupon2009},  have  managed  to  control  random  and
systematic  photometric  uncertainties, which  in  turn translates  to
large   improvements   in   photometric  redshift   estimates.    This
development has  motivated methods that use  photometric redshifts, in
combination  with   spectroscopy,  to  determine   the  clustering  of
extragalactic populations \citep {Myers2009, Hickox2011}.

In this paper we present a method similar to that of \cite {Myers2009}
that uses photometric redshifts only for galaxies and spectroscopy for
X-ray  AGN  to estimate  the  AGN/galaxy real-space  cross-correlation
function and infer the  bias and DMH mass of AGN. It  is shown that by
weighing  each  galaxy   with  its  photometric  redshift  Probability
Distribution  Function  yields  clustering  results similar  to  those
obtained  using   spectroscopy  for  both  AGN   and  galaxies.   This
methodology is then  applied to all X-ray fields  with public and good
quality photometric redshift  estimates and extensive spectroscopy for
X-ray sources.  This translates to a significant increase, compared to
previous  studies, in  the size  of the  X-ray AGN  sample  and better
control on the impact of sample variance on the results.

In Section 2 the AGN and galaxy samples are presented. The photometric
redshift  based methodology  for the  determination of  the AGN/galaxy
projected  real-space  cross-correlation   function  is  described  in
Section 3.  The  results are presented in Section  4 and are discussed
in     Section    5.      Throughout    this     paper     we    adopt
$H_0=100$\,km\,s$^{-1}$\,Mpc$^{-1}$,         $\Omega_m=0.3$        and
$\Omega_\Lambda=0.7$ and  $\sigma_8=0.8$. Rest frame  quantities (e.g.
luminosities,   dark   matter  halo   masses)   are  parametrised   by
$h=H_0/100$, unless otherwise stated.

\section{The data} \label{section:samples}

Three extragalactic survey fields are used to determine the clustering
of X-ray  AGN at $z\approx1$.   The Extended Chandra Deep  Field South
Survey (ECDFS), the All  Wavelength Extended Groth strip International
Survey  \citep[AEGIS,][]{Davis2007}  and  the  Cosmological  evolution
Survey  \citep[COSMOS,][]{Scoville2007}.   The  choice  of  fields  is
motivated  by  the availability  of  (i)  X-ray  data, (ii)  extensive
follow-up spectroscopic programs  targeting specifically X-ray sources
and (ii) deep multiwavelength  imaging (UV, optical, infrared) for the
determination   of  photometric   redshift   Probability  Distribution
Functions (PDFs) for individual galaxies.

\subsection{Optically selected galaxy sample} \label{section:galaxies}

The  COSMOS  and AEGIS  fields  overlap with  the  D2  and D3  regions
respectively,  of  the  deep synoptic  Canada-France-Hawaii  Telescope
Legacy Survey (CFHTLS).  The optical photometry ($ugriz$ bands) of the
T0004  data  release  is  used, which  includes  photometric  redshift
estimates reliable  to $i_{AB}<24$\,mag and  photometric redshift PDFs
\citep{Coupon2009}.    Regions   of   unreliable  photometry   (CFHTLS
catalogue   parameter  {\sc   flag\_terapix$>1$})  because   e.g.   of
contamination by  bright stars,  are masked out.   In the  analysis we
only  use  CFHTLS  optical  sources  classified  as  galaxies  (CFHTLS
parameters {\sc  object} and {\sc flag\_terapix} equal  to zero), with
reliable  photometric   redshift  estimates  (CFHTLS   parameter  {\sc
zp\_reliable}$\neq-99$)      and     $17.5<i_{AB}<24$\,mag.      Table
$\ref{table:gal_samples}$ lists  the total number of  galaxies used in
the D2 and D3 CFHTLS regions.

In  the  ECDFS  the   photometric  catalogue  compiled  by  the  MUSYC
collaboration  is used  \citep{Cardamone2010}.   They combined  Subaru
narrow-band imaging with existing $UBVRIzJHK$ photometry and $Spitzer$
IRAC images to create a  uniform catalogue of $\sim40,000$ galaxies to
$R_{AB}=25.3$.   Reliable photometric  redshifts, including  PDFs, are
also  estimated to  that magnitude  limit  \citep{Cardamone2010}.  Our
analysis  uses MUSYC  optical  sources classified  as galaxies  (MUSYC
catalogue parameter  {\sc star\_flag}$<1$; Cardamone et  al. 2010) and
magnitudes in the range $17.5<i_{AB}<24$\,mag, as for CFHTLS galaxies.
Table  $\ref{table:gal_samples}$ lists  the total  number  of galaxies
used in the MUSYC field.

\subsection{AGN samples}

X-ray AGN are selected from the AEGIS-800\,ks survey (AEGIS-XD, Nandra
et     al.      in     prep),     the    Chandra     COSMOS     survey
\citep[C-COSMOS,][]{Elvis2009} and  the ECDFS \citep{Lehmer2005}.  The
AEGIS-XD  covers a  total of  $\rm 0.3\,deg^2$  in the  Extended Groth
Strip  to  a  total  depth  of 800\,ks  by  combining  Chandra  ACIS-I
observations carried out in AO-3,  AO-6 and AO-9. The AEGIS-XD and the
ECDFS X-ray source catalogues  are generated following the methodology
of \cite{Laird2009}.   For C-COSMOS the source  catalogue presented by
\cite{Elvis2009}  is used.  In all  three  surveys we  use all  X-rays
sources independent of the energy band in which they are detected.

The  optical identification  of  the  X-ray sources  is  based on  the
Likelihood   Ratio   method   \citep{Sutherland_and_Saunders1992}   as
described in  \cite{Georgakakis2009}.  The AEGIS-XD  and COSMOS source
lists are matched to the T0004 release optical catalogue of the CFHTLS
\citep{Coupon2009}. The  ECDFS X-ray sources are  cross-matched with the
MUSYC photometric catalogue \citep{Cardamone2010}.

Optical spectroscopy of X-ray sources  in the AEGIS field is primarily
from the  DEEP2 \citep{Newman2012}  and DEEP3 galaxy  redshift surveys
\citep{Cooper2011_deep3_1, Cooper2012_deep3_2} as well as observations
carried  out  at  the  MMT  using  the  Hectospec  fibre  spectrograph
\citep{Coil2009}.   Spectroscopic  redshifts in  the  ECDFS have  been
compiled  by  \cite{Cardamone2010}.  We  also include  redshifts  from
\cite{Silverman2010} and  \cite{Cooper2011_aces}.  Redshifts in COSMOS
are  from the  public  releases of  the  VIMOS/zCOSMOS bright  project
\citep{Lilly2009}   and  the   Magellan/IMACS   observation  campaigns
\citep{Trump2009}, as  well as the compilation of  redshifts for X-ray
sources presented by \cite{Brusa2010}.

In  the  clustering  analysis  we  use  X-ray  sources  with  $L_X(\rm
2-10\,keV) > 10^{41} \, erg \, s^{-1}$ and redshifts $0.7<z<1.4$.  The
rest-frame  2-10\,keV   X-ray  luminosity  is   estimated  assuming  a
power-law spectral  energy distribution with  index $\Gamma=1.9$.  The
vast  majority   of  sources  above   this  luminosity  cut   are  AGN
\citep{Georgakakis2011}.   X-ray  sources in  regions  that have  been
masked out because of poor optical photometry (e.g.  bright stars) are
excluded from the analysis.  Table $\ref{table:agn_samples}$ shows for
each field the  number of X-ray AGN used  for clustering measurements,
their mean redshift and average X-ray luminosity.

\begin{table}
\caption{Galaxy  samples.  ``Full  sample" refers  to galaxies  in the
 optical catalogues  of the MUSYC,  CFHTLS-D2 and D3  after filtering 
out masked  regions and applying  the magnitude limits  discussed in
  the  text.  The  ``resampled"  galaxy  sample is  selected  to  have
  redshift   distribution  similar   to  that   of  X-ray   AGN.   The
  ``DEEP2-like" sample applies the photometric criteria adopted by the
  DEEP2 redshift survey team to select galaxies in
  the   redshift    range   0.7-1.4.    For    details   see   Section
  $\ref{section:auto-cor}$ and Appendix A.}\label{table:gal_samples}
\centering
\setlength{\tabcolsep}{2.5mm}
\begin{tabular}{cccc}
       \hline
{field} & {full sample} & {``resampled" } &{``DEEP2-like''} \\
 & & galaxies & galaxies \\
(1) & (2) & (3) & (4) \\
       \hline
CFHTLS (D2) & 55,367 & 28,150 & 21,731 \\
CFHTLS (D3) & 58,665 & 33,389 & 23,060 \\
MUSYC & 22,730 & 15,169 & 9,577 \\
       \hline
\end{tabular}
\end{table}

\begin{table}
\caption{X-ray AGN samples in the COSMOS, AEGIS-XD and ECDFS fields with $L_X(\rm
2-10\,keV) > 10^{41} \, erg \, s^{-1}$ and redshifts $0.7<z<1.4$.}
\centering
\setlength{\tabcolsep}{3.0mm}
\begin{tabular}{lccc}
       \hline
{field} & {No. of sources} & {$\rm <z>$} &{$\rm <\log L_X>$ } \\
 & & & ($\rm erg \,s^{-1}$) \\
       \hline
COSMOS & 182 & 0.97 & 43.3 \\
AEGIS-XD & 158 & 0.97 & 42.5 \\
ECDF-S & 101 & 1.02 & 42.8 \\
       \hline
\label{table:agn_samples}
\end{tabular}
\end{table}

\section{Methodology}\label{section:method}

\subsection{AGN/galaxy Cross-Correlation Function}

In real-space, the AGN/galaxy cross-correlation function, $\xi(r)$, is
given by

\begin{equation}
\label{eqn:xi} \xi(r) =\frac{DD(r)}{DR(r)}-1,
\end{equation}

\noindent  where DD(r)  and DR(r)  are the  AGN/galaxy  and AGN/random
pairs at separation $r$.  The random point catalogue should follow the
galaxy  sample  selection   function,  i.e.   magnitude  limit,  field
boundaries, masked regions.

The distance $r$ can be  decomposed into separations along the line of
sight, $\pi$,  and across  the line of  sight, $\sigma$. If  $s_1$ and
$s_2$  are   the  distances   of  two  objects   1,  2,   measured  in
redshift-space, and $\theta$ the angular separation between them, then
$\sigma$ and $\pi$ are defined as

\begin{equation} 
\pi=(s_2-s_1), $ along the line-of-sight$,
\end{equation}

\begin{equation}
\sigma=\frac{(s_2+s_1)}{2}\theta , $ across the line-of-sight$.
\end{equation}

\noindent The correlation function in redshift-space is then estimated as

 \begin{equation}
\xi(\sigma,\pi) =\frac{DD(\sigma,\pi)}{DR(\sigma,\pi)}-1.
\label{eqn:wtheta}
\end{equation} 

\noindent This simple estimator has the advantage that the
  random point source catalogue needs to account only for the
  selection function of galaxies, which is typically a spatial
  filter. In contrast more advanced clustering estimators, such as
  that of \cite{LZ1993}, require the construction of random
  catalogues for the X-ray source population as well.  This might introduce
  systematic biases into the calculations  because X-ray  observations
  have variable sensitivity accross the field of view, which is
  challenging to quantify  accurately.

\noindent  In  the  classic  approach  of  estimating  the  real-space
cross-correlation function, spectroscopic  redshifts are available for
the AGN and  galaxy samples.  For each AGN/galaxy  pair with $\sigma$,
$\pi$ separations  determined from their sky  positions and redshifts,
the     $DD(\sigma,\pi)$    is     incremented     by    one,     i.e.
$DD(\sigma,\pi)=DD(\sigma,\pi)+1$.  This  relation can be  modified to
account for uncertainties in the determination of the redshifts of the
galaxy population.  In  this case it is assumed  that the redshifts of
galaxies follow continuous  probability distribution functions
(i.e.  photometric redshifts), while those  of AGN are known at a high
level of accuracy (i.e.  spectroscopic redshifts).  For a given galaxy
the probability $f_{gal}$ that  it lies at separations $\sigma$, $\pi$
from an AGN  can be estimated from its  PDF.  $DD(\sigma,\pi)$ is then
incremented     by    $f_{gal}$,     instead     of    unity,     i.e.
$DD(\sigma,\pi)=DD(\sigma,\pi)+f_{gal}$.   Therefore,   a  galaxy  may
contribute with  different probabilities, $f_{gal}$, to  more than one
$(\sigma,\pi)$ bins of an AGN/galaxy pair.  $DR(\sigma,\pi)$ pairs are
estimated  following  the   same  procedure.   Random  catalogues  are
generated by randomising the positions of galaxies taking into account
the  geometry and  masked  regions  of each  optical  survey. In  this
approach  every  random point  has  attached  to  it the  photometric
redshift PDF  of the galaxy  from which it  was generated. A  total of
three random  catalogues are produced,  one per survey field.  Each of
them has the same size as  the real galaxy catalogue from which it was
generated.

When  the  correlation function  is  measured  in redshift-space,  the
clustering is affected at small  scales by the rms velocity dispersion
of AGN along the line of  sight and by dynamical infall of matter into
higher density regions.  To first  order, only the radial component of
$\xi(\sigma,\pi)$  is affected by  redshift-space distortions.  We can
therefore remove  this bias  by integrating along  the line  of sight,
$\pi$,   to  calculate   the   projected  cross-correlation   function

\begin{equation}
w_p(\sigma)=2\int_0^{\pi_{max}} \xi(\sigma,\pi)d\pi.
\label{eqn:wp}
\end{equation}

\begin{figure}
\begin{center}
\includegraphics[scale=0.28]{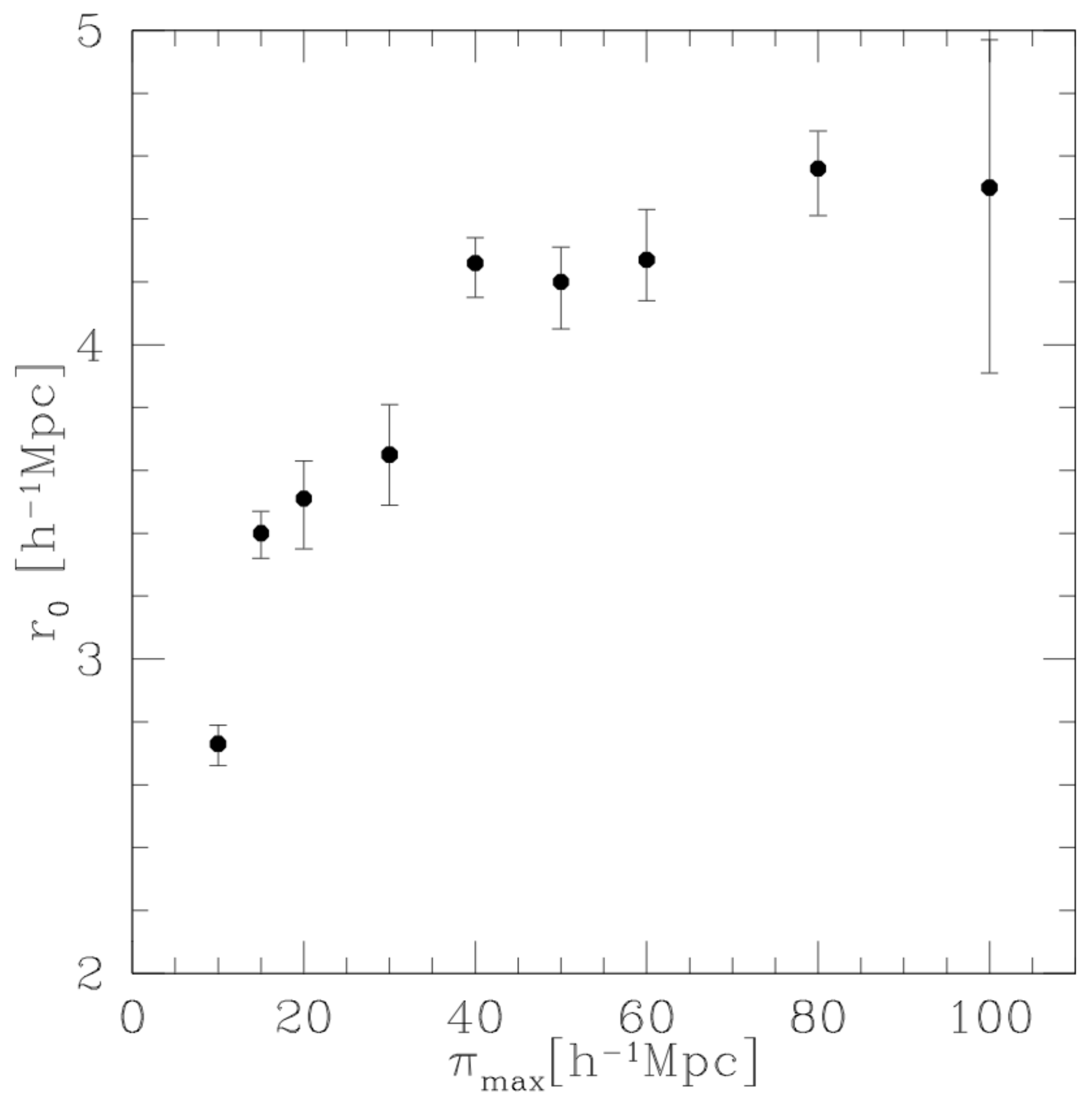}
\caption{The AGN/galaxy cross-correlation length, $r_0$, as a function
of  $\pi_{max}$, the  maximum  scale of  the  integration in  equation
\ref{eqn:wp}.  The cross-correlation function is measured by combining
the AEGIS, ECDFS and COMSOS fields. The errors are jackknife.}
\label{fig:pimax}
\end{center}
\end{figure}

\noindent The maximum scale of  the integration is a trade-off between
underestimating the clustering amplitude, if $\pi_{max}$ is too small,
and  low signal-to-noise  ratio,  if $\pi_{max}$  is  too large.   The
optimum  $\pi_{max}$ value  is determined  by measuring  the projected
AGN/galaxy cross-correlation function in the combined AEGIS, ECDFS and
COSMOS fields (see next section for details) for different $\pi_{max}$
values in the range $10-100h^{-1}$Mpc.  Figure \ref{fig:pimax} shows
that  the  amplitude  of  the  AGN/galaxy  cross-correlation  function
saturates for $\pi_{max}=40$\,Mpc.  This is  the value we adopt in the
analysis.

Assuming  that the  real-space  AGN-galaxy cross-correlation  function
follows  a  power-law  of  the  form  $\xi(r)=(r/r_o)^{-\gamma}$,  we estimate
the real-space   cross-correlation  amplitude,   $r_0$,  and   the  slope,
$\gamma$ from the relation

\begin{equation}
\frac{w_p(\sigma      )}{\sigma}=\left(\frac{r_0}{\sigma}\right)^\gamma
\frac{\Gamma                                        (\frac{1}{2})\Gamma
  (\frac{\gamma-1}{2})}{\Gamma(\frac{\gamma}{2})},
\label{eqn:projected_ro}
\end{equation}

\noindent where  $\Gamma (x)$ is  the Gamma function.   The AGN-galaxy
cross-clustering strength  can then be  expressed in terms of  the rms
fluctuation of the density distribution  over a sphere with a comoving
radius of 8 $h^{-1}$Mpc \citep[e.g.][] {Miyaji2007}

\begin{equation}
\sigma_{8,AG}^2=J_2(\gamma)\left(                \frac{r_0}{8h^{-1}Mpc}
\right)^\gamma,
\label{eqn:s8}
\end{equation}
where

\begin{equation}
J_2(\gamma)=\frac{72}{(3-\gamma)(4-\gamma)(6-\gamma)2^\gamma},
\end{equation}

\noindent the AGN/galaxy bias, $b_{AG}$, can then be calculated via

\begin{equation}
b_{AG}=\frac{\sigma_{8,AG}}{\sigma_8(z)}.
\label{eqn:bs8}
\end{equation}

\subsection{Galaxy auto-correlation function and AGN bias}\label{section:auto-cor}

An estimate of the galaxy bias  is required to infer the AGN bias from
the  AGN/galaxy   cross-correlation  function.   If   spectroscopy  is
available for galaxies their bias can be determined from the projected
real-space   galaxy   auto-correlation   function   (e.g.    equations
$\ref{eqn:s8}-\ref{eqn:bs8}$).

In  the absence  of  spectroscopy,  the galaxy  bias  is estimated  by
deprojecting to 3-D \citep {Limber1953} their angular auto-correlation
function, $w(\theta)$.  This calculation  assumes a power-law form for
the correlation function.  It  also requires knowledge of the redshift
distribution,  N(z),   of  the  galaxy  population.    The  latter  is
determined by  summing the photometric redshift PDFs  of galaxies.  In
section 4.2 we confirm that  the galaxy bias inferred from $w(\theta)$
is consistent with that  determined from 3-D clustering analysis using
large galaxy spectroscopic surveys like DEEP2.

\begin{figure}
\begin{center}
\includegraphics[scale=0.28]{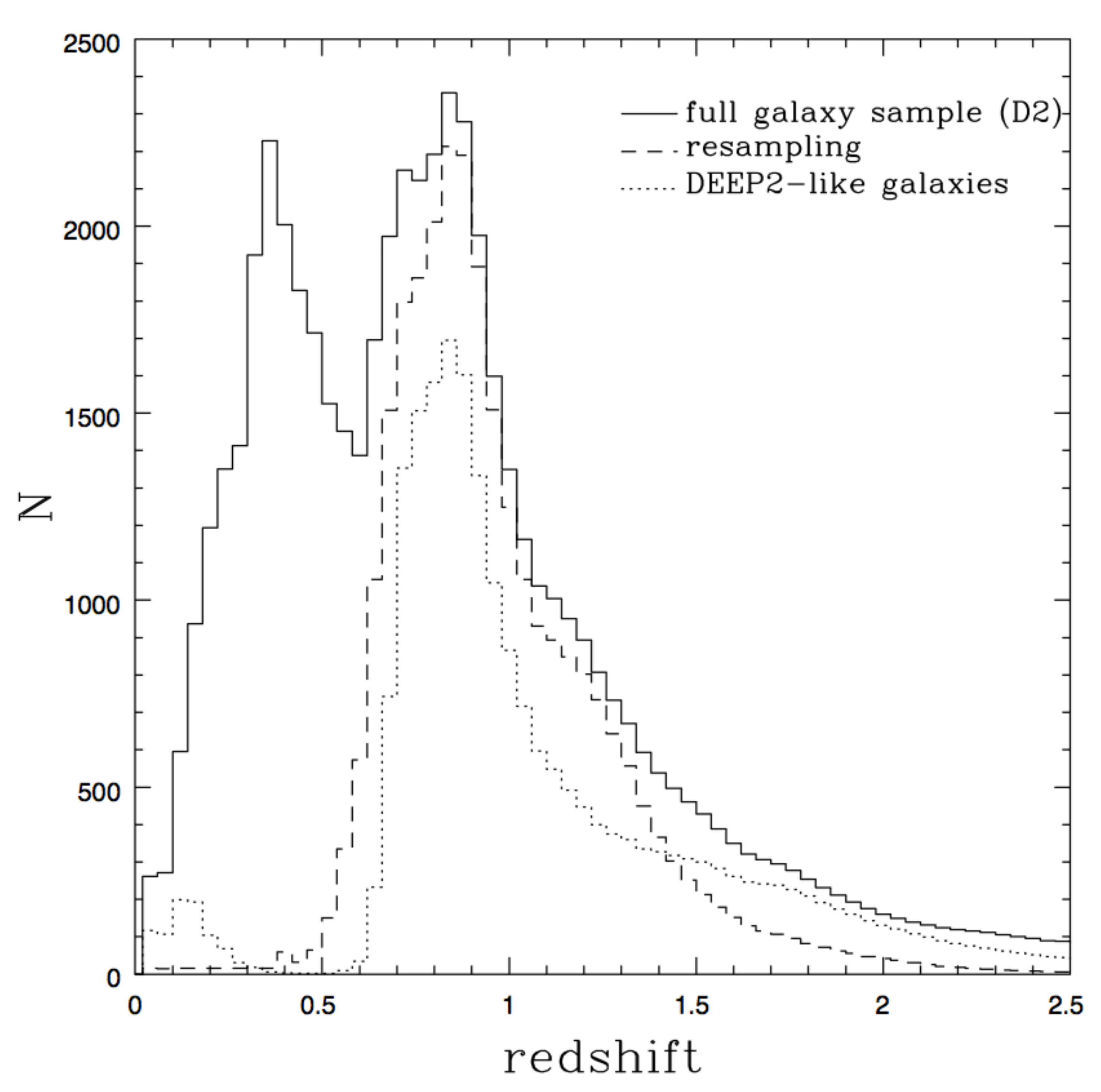}
\caption{Solid line shows the redshift distribution of the full CFHTLS
  galaxy sample (55,367) in the COSMOS field. Dashed line presents the
  N(z) of the ``resampling method" galaxy sample (total of 28,150; see
  Section  $\ref{section:auto-cor})$. The  N(z) of  the ``DEEP2-like''
  galaxies  (21,731;  see  Appendix  A)  is plotted  with  the  dotted
  line. The N(z) of each galaxy population is estimated by summing the
  photometric redshift PDFs of individual sources.}
\label{fig:gal_nz}
\end{center}
\end{figure}

Care should  be exercised in the  selection of the  galaxy sample for
which  the bias  is estimated  to infer  the AGN  clustering  from the
cross-correlation function.   Fig.  $\ref{fig:gal_nz}$ shows  that the
redshift  distribution of  the overall  galaxy population  at  a given
magnitude limit is wide and  includes a large fraction of low redshift
sources, outside the  redshift interval of the X-ray  AGN sample (i.e.
0.7--1.4).  One  expects that the  AGN/galaxy cross-correlation signal
is  dominated by  galaxies with  photometric redshift  PDFs  that peak
within the redshift interval of the  AGN population. It is the bias of
those  galaxies  that  should  be  used to  infer  the  clustering  of
AGN.   Using the  full galaxy  sample at  a given  magnitude  limit to
determine the  AGN bias yields erroneous results.   Two approaches are
adopted to account for this effect. 

The  first method  limits the  galaxy sample  by  applying appropriate
optical colour selection criteria to exclude from the analysis sources
outside the redshift range of interest.  We adopt the colour selection
used by  the DEEP2 team to  pre-select galaxies in  the redshift range
0.7-1.4  \citep{Newman2012}, i.e  at the  same redshift  range  as the
X-ray AGN.  Appendix A describes how the DEEP2 colour pre-selection is
applied to the CFHTLS and MUSYC photometric catalogues.  The resulting
galaxy samples are referred to  as ``DEEP2-like''. The total number of
galaxies in those samples are listed in Table \ref{table:gal_samples}.
Figure $\ref{fig:gal_nz}$ plots the  N(z) of the ``DEEP2-like'' galaxy
sample in the  CFHTLS-D2 field and demonstrates the  efficiency of the
DEEP2 colour cuts in eliminating  sources at $z<0.7$.  The size of the
``DEEP2-like'' galaxy samples in the  CFHTLS and MUSYC survey areas is
presented  in Table  \ref{table:gal_samples}.  This  approach  has the
advantage that one has some  control on the properties of the galaxies
that go into the cross-correlation function.

The  alternative  approach  is  similar  to that  presented  by  \cite
{Hickox2011}.  Assume  a particular AGN $i$  at spectroscopic redshift
$z_i$ and  a galaxy $j$ which contributes  $f_{gal,i,j}$ to AGN/galaxy
pairs at a particular scale $DD(\sigma, \pi)$.  For the galaxy $i$ one
can estimate the  weight $w_j=\sum_{i} f_{gal, i, j}$,  the sum of its
contribution  to all  pairs with  all  the AGN  in the  sample at  all
scales.  The quantity  $w_j$ defines the average probability that
a  particular galaxy participates  in the  cross-correlation function.
One  can  weigh each  galaxy  with  $w_j$  to construct  the  redshift
distribution of the galaxies  that contribute to the cross-correlation
signal.  One can then randomly  draw samples from the galaxy catalogue
that follow  the weighted N(z).  The  bias of those  galaxies is then
used to infer the AGN clustering from the AGN/galaxy cross-correlation
function.  The  galaxy samples produced by this  approach are referred
to  as the  ``resampling method"  samples.   Their size  in the  three
survey  fields  is  shown  in  Table  $\ref{table:gal_samples}$.   The
redshift distribution of the ``resampling method" galaxy sample in the
CFHTLS-D2 field is presented in Fig. $\ref{fig:gal_nz}$.

Equations  $\ref{eqn:s8}-\ref{eqn:bs8}$ are used  to measure the
AGN/galaxy bias, $b_{AG}$,  and the galaxy bias, $b_g$.  The X-ray AGN
bias, $b_{AGN}$, is

\begin{equation} 
b_{AGN}=\frac{b^2_{AG}}{b_g}.
\end{equation} 

\noindent  The  AGN DMH  mass  is inferred  from  the  bias using  the
methodology   of  \cite[e.g.][]{Mo1996}.    In  brief,   assuming  the
ellipsoidal collapse  model of \cite  {Sheth2001} the bias  values are
converted  to $\nu=\delta_c/\sigma(M)$   \cite[equation (23)  of
][]{daAngela2008},  where $\sigma(M)$  is the  rms fluctuation  of the
density  field  and $\delta_c\approx  1.69$  the critical  overdensity
required  for collapse.  Then $\nu$  is  converted to  DMH mass  using
equations (A8)-(A10) of \cite {Bosch2002}.

\subsection{Combining clustering results from different fields} 

The AGN-galaxy cross-correlation  function and galaxy auto-correlation
function  are determined  from the  combined AEGIS,  COSMOS  and ECDFS
fields. This is to improve  the statistical reliability of the results
and  minimise   the  impact  of  cosmic-variance   on  the  clustering
measurements.  The combined correlation function is

\begin{equation}
 \xi_{all}(r) =\frac{DD_{all}(r)}{DR_{all}(r)}-1,
\end{equation} 

\noindent where

\begin{equation} 
DD_{all}(r)=\sum_{i=1}^3 DD_i(r).
\end{equation} 

\noindent  $DD_{i}(r)$  is  AGN-galaxy  pairs  in each  of  the  three
fields. The same relation is used in the case of DR pairs.

The  uncertainties in the  correlation function at a  given scale
  are estimated  using the Jackknife methodology. Each  field is split
  into  five equal-size  subregions. This  results  to a  total of  15
  sections spread  over the three  fields of choice.   The correlation
  function is  determined 15 times,  excluding each time  one section.
  The jackknife error is 

\begin{equation}
\sigma ^2=\frac{N-1}{N}\sum_{L=1}^{N}\frac{DR_L}{DR}[\xi_L-\xi]^2,
\end {equation}

\noindent where  N is the number  of sections, i.e. 15,  $DR_L$ is the
data-random  pairs  in each  section,  $DR$  is  the total  number  of
data-random  pairs, $\xi_L$  is the  correlation function  measured in
each section and  $\xi$ is the combined correlation  function. We
  experimented with different number  of Jackknife regions. For a very
  small  number of  regions one  expects errors  dominated  by poisson
  noise.  Similarly,  for very large  number of subregions  the sample
  variance  uncertainties become  negligible and  shot noise  is again
  important.  The estimated Jackknife errors normalised to the Poisson
  errors (excess  variance) at a given scale  increases initially with
  the  number  of  Jackknife  regions,  reaches  a  plateau  and  then
  decreases. The  chosen value of  15 Jackknife regions is  within the
  plateau.

\begin{figure}
\begin{center}
\includegraphics[scale=0.28]{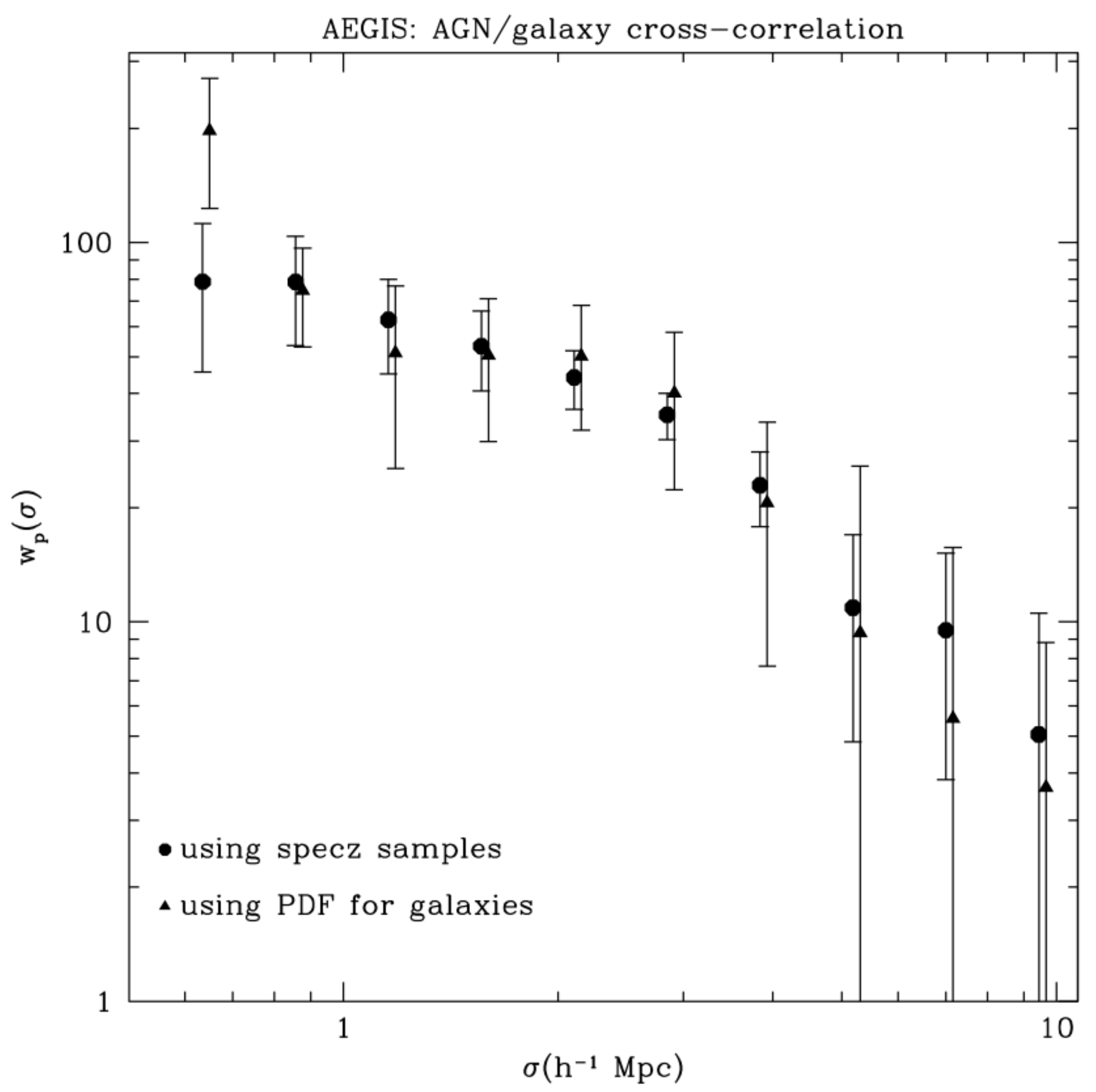}
\caption{The   projected   real-space   AGN/galaxy   cross-correlation
  function in  the AEGIS  field.  Triangles are  the cross-correlation
  function using photometric redshifts  only for the galaxy population
  (CFHTLS-D3). Circles correspond to the cross-correlation function using
  only galaxies with spectroscopic  redshifts from the DEEP2 and DEEP3
  surveys. Errors are jackknife.  For clarity, triangles are offset in
the horizontal direction by $\delta$log$\sigma$=+0.01.}
\label{fig:specz_vs_photoz}
\end{center}
\end{figure}

\section{Results} 

\subsection{Performance of the photometric redshift based AGN/galaxy cross-correlation method}

Before applying  the methodology presented in the  previous section to
the combined AEGIS, ECDFS and COSMOS fields we verify that it produces
similar results with the traditional approach which uses spectroscopic
redshift for both the AGN  and the galaxy population. In this exercise
we use data  from the AEGIS field only,  which benefits from extensive
spectroscopy  for galaxies  as part  of  the DEEP2  and DEEP3  surveys
\citep{Cooper2011_deep3_1, Cooper2012_deep3_2, Newman2012}.

Fig.    $\ref{fig:specz_vs_photoz}$  presents   the  cross-correlation
function of  158 X-ray AGN  with $\sim 5,100$ spectroscopic  DEEP2 and
DEEP3 galaxies with $R<24.1$\,mag. This is compared with the projected
cross-correlation  function estimated  by replacing  the spectroscopic
galaxy sample with  $\sim 23,000$ photometric CFHTLS-D3 ``DEEP2-like''
galaxies.  The estimated  AGN bias is $b_{A}=1.93^{+0.40}_{-0.37}$ and
$b_{A}=2.05^{+0.21}_{-0.20}$, for  the photo-z based  and the standard
method,   respectively  (power   law  fits   are  applied   on  scales
1-10$h^{-1}$Mpc).   The two  methods yield  consistent results  on the
inferred bias of AGN.

\subsection{Determining the bias and DMH mass of X-ray AGN at z=1}

\begin{figure}
\begin{center}
\includegraphics[scale=0.28]{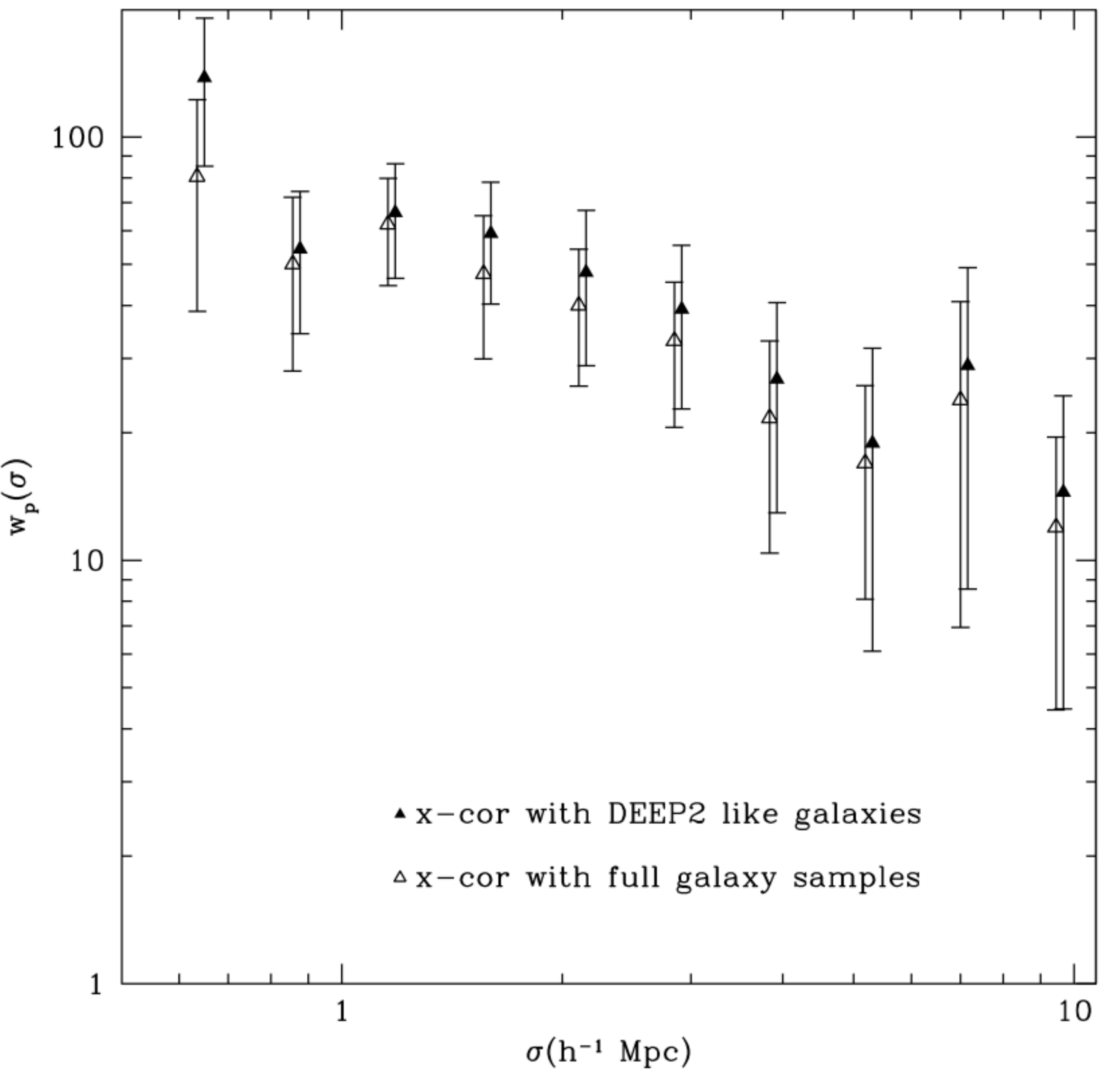}
\caption{AGN-galaxy  projected  cross-correlation  function  from  the
  combined AEGIS,  COSMOS and  ECDFS fields.  Filled  triangles present
  the results  using the ``DEEP2-like'' galaxy  sample. Open triangles
  show the results using the full galaxy samples.  For clarity, filled
  triangles    are   offset   in    the   horizontal    direction   by
$\delta$log$\sigma$=+0.01.}
\label{fig:wp_combined}
\end{center}
\end{figure}

Fig.    $\ref{fig:wp_combined}$  presents  the   projected  AGN-galaxy
cross-correlation function  from the combined AEGIS,  ECDFS and COSMOS
fields.   Two sets  of points  are plotted.  One corresponds  to the
cross-correlation  function with all  galaxies and  the other  to the
cross-correlation  with the  ``DEEP2-like'' galaxy  samples.   The two
cross-correlation  functions are not  directly comparable  because the
galaxy samples used in the calculations are different.

Assuming  a  power-law  form  for the  cross-correlation  function  we use a $\chi^2$ minimization procedure to
estimate the  best-fit amplitude, $r_0$, and  the slope, $\gamma$,
at  scales   $1-10h^{-1}$\,Mpc,  in the linear regime of clustering. The  results  are   shown  in  Table
\ref{table:fits}.  The fits use only the diagonal elements of the covariance matrix. This
does not affect the best fit values, but may bias the inferred minimum
reduced $\chi^2$ to values lower than unity. The  errors correspond  to  the 68th  percentile
around the  minimum $\chi^2$. Non-linearities may  affect the correlation function on scales larger than $1h^{-1}$\,Mpc \citep{RossA2011}. Changing our fitting scales to $0.5-10h^{-1}$\,Mpc and $3-10h^{-1}$\,Mpc, the inferred AGN bias is stable at the 10\% level  consistent  within the errors  with the values  estimated at scales 1-10Mpc.

Table  \ref{table:fits}, also presents
the  AGN-galaxy bias, $b_{AG}$,  which has  been calculated  using the
best-fit  $r_0$, $\gamma$  and  equations $\ref{eqn:s8}-\ref{eqn:bs8}$.
The effective redshift distribution for the cross-correlation function
in each  of the three  fields is estimated  by the convolution  of the
redshift   distributions  of   the  AGN   and  galaxy   samples,  i.e.
$N_{AG}(z)=N_{AGN}(z)\times   N_{gal}(z)$.    The   overall   redshift
distribution     is     calculated     in    a     similar     manner,
$N^{all}_{AG}(z)=N^{AEGIS}_{AG}(z)\times       N^{COSMOS}_{AG}(z)\times
N^{ECDFS}_{AG}(z)$.

\begin{figure}
\begin{center}
\includegraphics[scale=0.28]{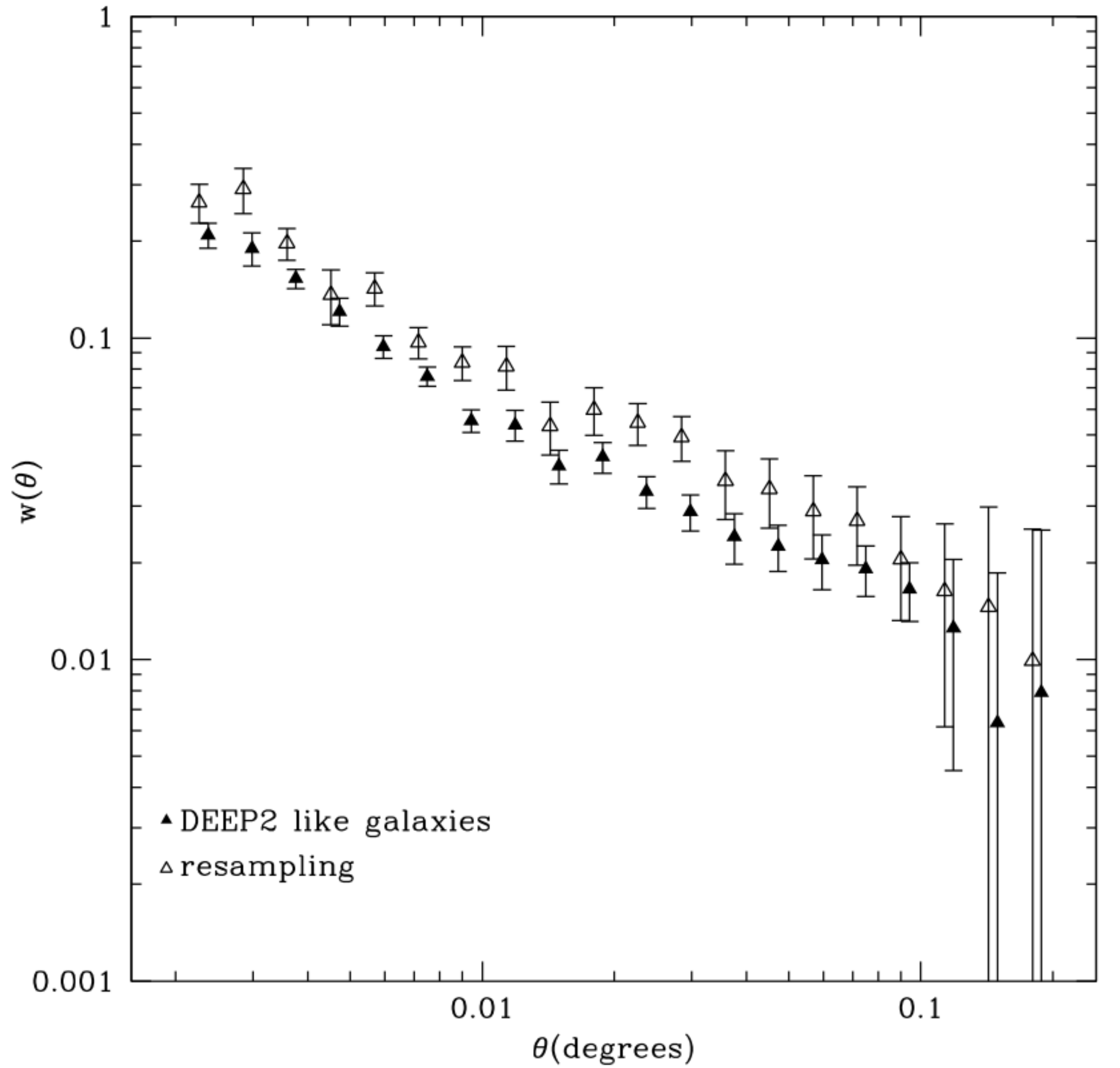}
\caption{Galaxy  angular  auto-correlation   function   from  the
  combined AEGIS,  COSMOS and ECDFS  fields.  Filled triangles  are the
  results using  the ``DEEP2-like'' galaxy sample.  Open triangles show
  the  results using  the  ``resampling method''  galaxy sample.  Open
  triangles are offset by -0.02 in the horizontal axis for clarity.}
\label{fig:wtheta_combined}
\end{center}
\end{figure}

Figure $\ref {fig:wtheta_combined}$  presents the measurements for the
galaxy  angular   auto-correlation  function for  the
``resampling method''  galaxy sample and  the ``DEEP2-like'' galaxies.
In fields of  finite size the angular auto-correlation function
estimation is biased to lower
values (integral constraint, \citealp {Groth1977}). The effect of this
bias  becomes  important  on  scales  comparable to  the  survey  size
($\sim0.5-1$  degrees  in  our   case).   Scales  in  the  range
  $0.025^\circ\leq \theta  \leq 0.2^\circ$ are used to  fit the galaxy
  auto-correlation function.  At  these scales the integral constraint
  is negligible.   Also, these angular  separations correspond
  to about $1-10\,$ h$^{-1}$Mpc  at $z\approx1$, i.e. the scales
  over which  the AGN/galaxy cross-correlation  function is measured.
Following  the procedure  described in  Section \ref{section:auto-cor}
the  galaxy  bias is  estimated.   The  results  are listed  in  Table
\ref{table:fits}.

As  a  check,  we  use  spectroscopically confirmed  galaxies  in  the
AEGIS-XD field  (DEEP2 and DEEP3  surveys) to determine  the projected
real-space   auto-correlation  function.    This   calculation  yields
$b_g=1.78^{+0.12}_{-0.11}$   consistent   with   the   bias   of   the
``DEEP2-like''    galaxy   sample    inferred    from   the    angular
auto-correlation   function,   $b_g=1.62\pm0.08$   (see   Table   \ref
{table:fits}).

Combining the AGN/galaxy projected cross-correlation function with the
bias of galaxies inferred from their angular auto-correlation function
yields  the   AGN  bias   and  DMH  masses   listed  in   Table  $\ref
{table:fits}$.  The cross-correlation functions using the ``resampling
method'' and the ``DEEP2-like'' galaxy samples yield $b_{AGN}$ and DMH
masses that are consistent within the statistical uncertainties.

\subsection {The effect of groups in the clustering signal}

In addition to the average dark matter halo mass of AGN, it is also
desirable  to   have  information   on  their  distribution   in  halo
mass. \cite{Georgakakis2008}  for example found that at $z\approx1$
the fraction of X-ay AGN relative to galaxies is similar in groups and
in the field, suggesting  diverse environments for the AGN population.
Recent results on the halo occupation of moderate luminosity X-ray AGN
are also consistent with a wide range of halo masses for these sources
\citep{Miyaji2011, Allevato2012}.   The signal  to noise ratio  of our
clustering signal  is insufficient  for halo occupation  analysis.  We
can  nevertheless  get  a  handle   on  the  dark  matter  halos  mass
distribution of  X-ray AGN by investigating how  the clustering signal
changes once sources in the most massive dark matter haloes within the
surveyed area, i.e. groups, are removed.

Groups are identified via their  diffuse X-ray emission in the Chandra
data  of  the  ECDFS  (Finoguenov   et  al.   in  prep)  and  AEGIS-XD
(Erfanianfar in  prep). In the case  of C-COSMOS both  the Chandra and
the XMM observations \citep{Cappelluti2009}  in that field are used to
identify  groups \citep{Leauthaud2010}.   The adopted  methodology for
detecting diffuse X-ray  sources is described in \cite{Finoguenov2007}
and  \cite{Leauthaud2010}.  Group  optical counterparts  and redshifts
are  estimated  by  searching  for red-sequence  galaxy  overdensities
within 0.5\,Mpc (physical) off the X-ray centroid.  The identification
of  group members  uses  the Bayesian  approach of  \cite{George2011},
which estimates the probability that a galaxy belongs to a group given
their  projected separation,  the redshifts  of the  galaxy  and group
(including  photometric  redshift   information)  and  the  background
density  of field  galaxies.   The groups  identified  in the  COSMOS,
AEGIS-XD and ECDFS have masses $\ga \rm 2 \times10^{13}\,M_{\odot}$ in
the redshift range 0.7-1.4.  The  X-ray group catalogue is by no means
complete.   It  is  affected   for  example,  by  the  variable  X-ray
sensitivity  across the  surveyed area,  the different  depths  of the
three X-ray  fields used in the  analysis and scatter  in the relation
between  DMH mass  and  X-ray luminosity.   Nevertheless, using  X-ray
selected groups allows us to  explore, to the first approximation, the
impact of the most massive  structures within the surveyed area on the
AGN clustering signal.

A total  of 22 X-ray  AGN in the  sample (5\%, 22/441)  are associated
with  group members,  6  AGN in  the ECDFS,  7  in C-COSMOS  and 9  in
AEGIS-XD. We re-estimate the cross-correlation function after removing
this small number  of sources from the X-ray AGN  catalogue as well as
optically selected  galaxies within 0.5\,Mpc from the  X-ray centre of
groups.   This has  a strong  impact on  the inferred  DMH  mass.  The
results  are   shown  in  Table  \ref{table:fits}.   A   DMH  mass  of
$12.68^{+0.27}_{-0.40}$  is estimated, about  0.5\,dex lower  than the
full AGN sample.  The bulk of the X-ray AGN population at $z\approx1$
lives in  moderate size dark  matter haloes.  It should  be emphasised
that the observed decrease in  the clustering signal is related to the
way  AGN  are  distributed  in  halo  mass.   It  is  not  because  of
serendipitous massive  structures within  the surveyed area  that bias
the inferred DMH mass to high values.  The impact of the latter effect
on our results is small because of the combination of different fields
and the cross-correlation approach we adopt.

\subsection {Clustering dependence on luminosity}

Finally  we investigate  the dependence  of the  clustering  signal on
X-ray  luminosity, $L_X$.  To  minimise possible  aliases between  the
redshift and luminosity we first split the AGN sample at the
median redshift, $z=0.97$.  The X-ray sources in each redshift bin are
further separated into two nearly equal size low and high X-ray luminosity
subsamples.  They are presented in Table $\ref{table:lum}$.  Because
the number of  sources in each subsample is low, we explore differences
in  the  clustering   using the  relative  bias 

\begin{equation}
b_{rel}=\sqrt{{w_p(\sigma)_{high-L_X}}/{w_p(\sigma)_{low-L_X}}}
\end{equation}

\noindent where  $w_p(\sigma)_{high-L_X}$, $w_p(\sigma)_{low-L_X}$ are
the  projected  cross-correlation  functions  for  the  high  and  low
luminosity subsamples, respectively. In this calculation we use scales 
in the range $1-10$h$^{-1}$Mpc.  The low-z and high-z subsamples yield
$b_{rel}=1.25\pm0.23$  and  $1.52\pm0.35$, respectively.   The
uncertainties are too large to comment on the luminosity dependence of
the AGN clustering in the present sample.

\begin{table*}
\caption{Power-law  best-fit   parameters  $r_0$,  $\gamma$   for  the
AGN/galaxy    cross-correlation    function    and   galaxy    angular
auto-correlation  function  using  the  methods discussed  in  Section
$\ref{section:auto-cor}$. The $\chi^2$ fits are using only the diagonal elements of the covariance matrix. $\chi^2_{\nu}$ present the reduced $\chi^2$ values. The inferred biases  are also listed. In the
case of the cross-correlation function  the fit is performed at scales
$1\leq r\leq  10 h^{-1}$Mpc. For the  galaxy auto-correlation function
the fit  is on scales $0.025^{\circ} \leq \theta  \leq 0.2^{\circ}$.  The
inferred  AGN  bias  and  DMH  mass  are  also  presented.}   \centering
\setlength{\tabcolsep}{0.45mm} \setlength{\extrarowheight}{1.5mm}
\begin{tabular}{lcccccccccc}
      \hline
  & $r_0$          & $\gamma$ & $\chi^2_{\nu}$ & $b_{AG}$  & $r_0$   & $\gamma$ & $\chi^2_{\nu}$ & $b_{g}$ & $b_{AGN}$ & $\rm \log M_{DMH}$ \\
  &($h^{-1}$\,Mpc) &          &     (dof)    &  &($h^{-1}$\,Mpc) &          &   (dof) &     &      & ($h^{-1}$\,M$_\odot)$    \\
	\hline
Resampled galaxies & $5.15^{+0.40}_{-0.44}$ & $1.80^{+0.21}_{-0.18}$ & 0.34 (7) & $1.85^{+0.13}_{-0.14}$ & $4.9\pm0.3$ & $1.70\pm0.06$ & 2.05  (8) & $1.75\pm0.08$ & $1.96^{+0.29}_{-0.31}$ & $12.91^{+0.22}_{-0.31}$ \\

DEEP2-like galaxies &$5.60^{+0.48}_{-0.50}$ & $1.72^{+0.25}_{-0.18}$ & 0.22 (7) & $1.96\pm0.15$ & $4.5\pm0.3$ & $1.60\pm 0.08$ &  0.96 (8) & $1.62\pm 0.08$ &  $2.38\pm0.38$ & $13.20^{+0.20}_{-0.27}$ \\

AGN in X-ray groups excluded & $4.46^{+0.43}_{-0.46}$ &
$1.72^{+0.29}_{-0.33}$ & 0.17 (7)  & $1.61^{+0.13}_{-0.14}$ & $4.3\pm0.3$ & $1.55\pm0.07$& 0.87 (8) &$1.62\pm0.08$ & $1.71^{+0.29}_{-0.31}$ & $ 12.68^{+0.27}_{-0.40}$ \\
       \hline
\label{table:fits}
\end{tabular}
\end{table*}

\begin{table*}
\caption{The number of X-ray AGN in $z-L_X$ subsamples. }  
\centering
\setlength{\tabcolsep}{1.8mm}
\begin{tabular}{lcccccccc}
       \hline
 Field      & $0.7<z<0.97$ & $0.7<z<0.97$ & $0.97<z<1.4$ & $0.97<z<1.4$ \\ 
       & $41.0<L_X<42.7$ & $L_X>42.7$ & $41.0<L_X<43.2$ & $L_X>43.2$ \\ 
		\hline
COSMOS & 25 & 69 & 22& 66 \\
AEGIS & 58 & 23 & 56 & 21 \\
ECDFS & 28 & 17 &32 &24\\
		\hline
		\hline
Total & 111 & 109 & 110 & 111 \\
	\hline
\label{table:lum}
\end{tabular}
\end{table*}

\begin{table*}
\caption{Clustering  measurements for  X-ray AGN from the literature.
  Columns are:  (1) reference to the AGN sample;  (2) name of the
  survey that the AGN sample was selected from; (3) number of sources
  used; (4) median redshift of the sample; (5) median X-ray luminosity
  of the sample; (6) the AGN bias  values  re-calculated using  the
  real-space power law fits quoted in the studies and equations
  $\ref{eqn:s8}-\ref{eqn:bs8}$.}        
\centering
\setlength{\tabcolsep}{1.2mm}
\scalebox{1.0}{
\begin{tabular}{cccccc}

       \hline  
       Study  & Sample  & no  &  z &  $L_X$ & b \\ 
	&  &  of objects& & (erg\,$s^{-1}$)  &    \\
        (1) & (2) & (3) & (4) & (5) & (6)   \\
       \hline 
       This  Work  (DEEP2-like) & COSMOS/AEGIS-XD/ECDFS  &  441  &  0.97 &  42.9 &   $2.38\pm0.38$   \\ 

       This Work  (resampling) & COSMOS/AEGIS-XD/ECDFS  &  441  &  0.97  &  42.9 &   $1.96^{+0.29}_{-0.31}$  \\ 

	Mountrichas  \& Georgakakis (2012) & XMM/SDSS       &  297 & 0.10 & 42.1 & $1.23^{+0.12}_{-0.17}$ \\

       Coil et al. (2009)           &  AEGIS                                  &  113  &  0.90  &  43.2 &   $1.97\pm0.25$   \\ 

       Gilli et  al. (2005)          &  CDFN                                  &  160  &  0.96  &  43.3 &   $1.87^{+0.14}_{-0.16}$     \\  

      Gilli et al. (2005)           &  CDFS                                   &    97  & 0.84   &    43.5 &   $2.64\pm0.30$           \\ 

       Gilli  et  al.  (2009)       & COSMOS                              & 538   & 0.98   &     43.4 &   $3.08^{+0.14}_{-0.16}$ \\ 

	Yang et al. (2006)       & CDFN				          & 252   &  0.8     &     42.6  &   $1.77^{+0.20}_{-0.30}$\\ 

       Yang et al. (2006)       & CLASXS			         &  233  &   1.2    &      43.8 &    $3.58^{+0.85}_{-1.11}$\\ 

      Starikova  et al. (2011) & Bootes  			        &  1282$^*$ &   0.37  &  42.7  &   $1.55^{+0.15}_{-0.15}$ \\  

      Starikova  et al. (2011) & Bootes                             &  1282$^*$  &   0.74 &    43.4 &   $2.58^{+0.31}_{-0.31}$  \\  

     Starikova  et al. (2011) & Bootes                             &  1282$^*$  &   1.28 &    44.0 &   $2.93^{+0.43}_{-0.43}$  \\  

    Cappelluti et al. (2010)  &   BAT   			    &   199      &     0.045 &  43.5 &   $1.22^{+0.09}_{-0.08}$\\ 

   Allevato et al. (2011)      & COSMOS                        &   593      &    1.22  &           &   $2.80^{+0.22}_{-0.90}$  \\

   Krumpe et al.   (2010)   &  RASS                             &  1552     &  0.27   & 43.4     &  $1.11^{+0.10}_{-0.12}$\\

    Krumpe et al. (2012)   & RASS 				   & 629       &  0.13    & 42.8    &  $1.19^{+0.08}_{-0.09}$ \\

    Krumpe et al. (2012)   & RASS 				   & 1552       &  0.25    & 43.4    &  $1.06^{+0.09}_{-0.11}$ \\

    Krumpe et al. (2012)   & RASS 				   & 876       &  0.42    & 43.8    &  $0.96^{+0.22}_{-0.54}$ \\

\hline
\label{table:studies}
\end{tabular}}
\begin{list}{}{}
\item 
$^*$This number corresponds to the total number of sources in redshift range $0.17<z<4.5$. 
\end{list}
\end{table*}

\begin{figure*}
\begin{center}
\includegraphics[scale=0.6]{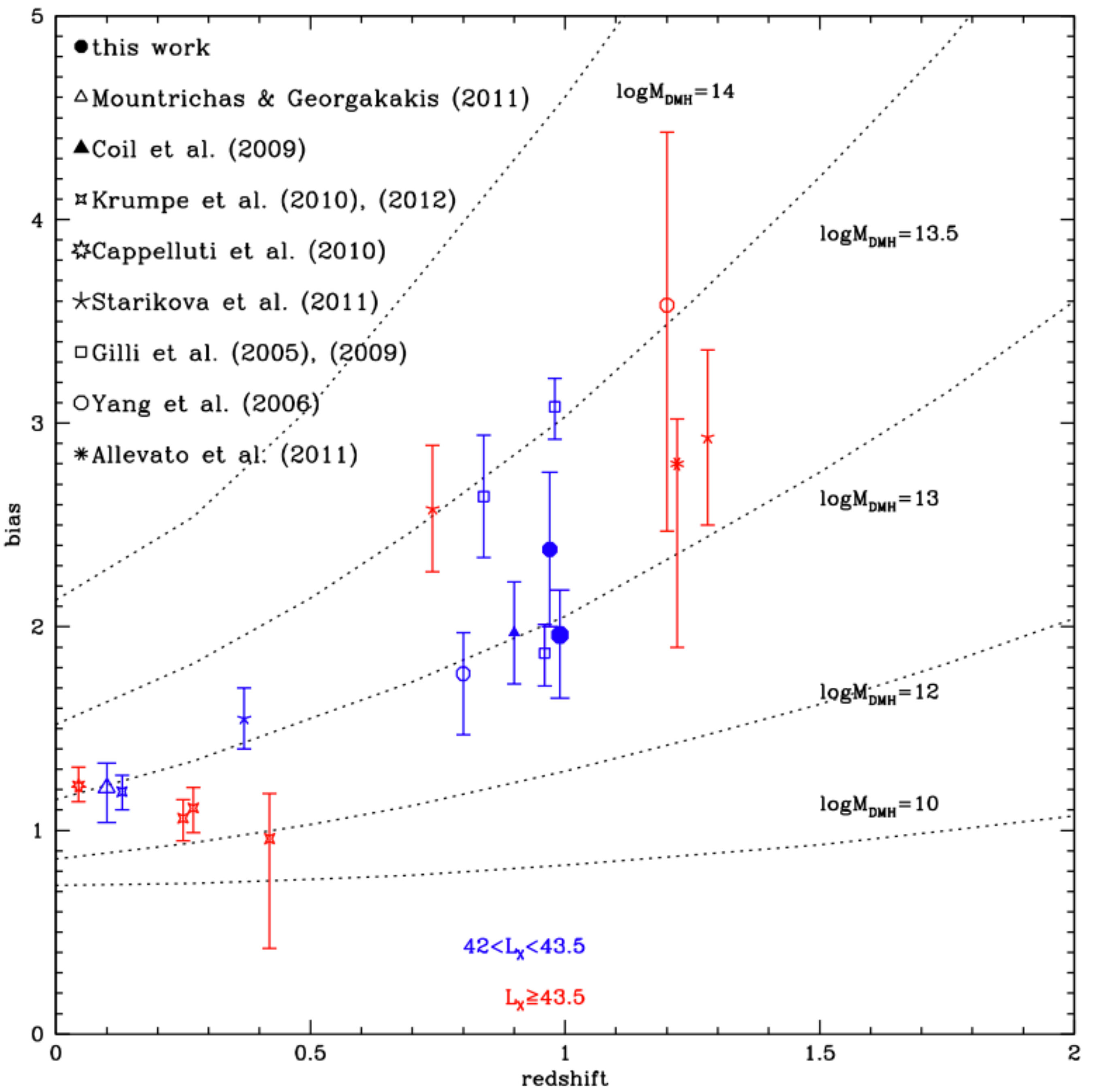}
\caption{The  inferred bias  of AGN  as a  function of  redshift.  Our
  measurements  using  both the  ``DEEP2-like''  and the  ``resampling
  method'' galaxy samples,  are plotted with the filled circles.   The AGN bias
  estimation  based  on the  ``resampling  method''  galaxy sample  is
  offset in the  horizontal axis by $\delta z=+0.02$  for clarity. The
  points have been  colour coded based on the  median X-ray luminosity
  of each  sample.  The  dotted lines present  the expected  $b(z)$ of
  DMHs.}
\label{fig:dmh}
\end{center}
\end{figure*}

\section{Discussion}

\subsection{The clustering estimation method}

A   method  is   presented  to   estimate  the   projected  real-space
cross-correlation  function  between   X-ray  AGN  and  galaxies  that
requires spectroscopy only for the AGN population and uses photometric
redshift probability distribution functions  for galaxies. The bias of
AGN is inferred by estimating  the clustering properties of the tracer
population, in our case  galaxies, from their angular auto-correlation
function and  (photometric) redshift distribution,  N(z).  A potential
source of systematic error at  this step is that the tracer population
may have a very different N(z) from the sources of interest, for which
one wishes  to estimate the  bias. In our  case, galaxies have  a much
wider N(z), that includes many low redshift sources, compared to X-ray
AGN.   \cite{Hickox2011} propose  the resampling  approach,  also used
here, to  account for this effect.  A downside of  this methodology is
that there  is little  control on the  selection of the  galaxies that
enter the  auto-correlation function determination.   These are chosen
in a  probabilistic manner  based on how  much they contribute  to the
cross-correlation clustering signal with AGN.

The resampling approach, and any uncertainties that this may introduce
in the results, can be avoided, if the galaxies are selected to have a
relatively narrow N(z)  that matches that of X-ray  AGN. An example of
this  approach  is  the  application  of  the  DEEP2  colour  cuts  to
photometrically pre-select galaxies  in the redshift interval $\approx
0.7-1.4$.  Resampling is not needed in this case.  We demonstrate this
point  by re-estimating  the AGN  bias after  applying  the resampling
methodology   to   the   ``DEEP2-like''   galaxy  samples   of   Table
\ref{table:gal_samples}.   This   exercise  yields  an   AGN  bias  of
$b_A=2.23^{+0.48}_{-0.50}$, in good  agreement with the bias estimated
using  the full  DEEP2-like  samples, $b_A=2.38\pm0.38$  (see
Table $\ref{table:fits}$).

Next we use LRGs (Luminous Red Galaxies) to further demonstrate that a
tracer population  with a narrow redshift distribution  is well suited
to the  clustering estimation  method based on  photometric redshifts.
Because of their  optical SEDs which is dominated by  old stars and is
characterised  by   a  prominent   4000\,\AA\,  break,  LRGs   can  be
photometrically selected in well  defined and relatively thin redshift
slices \citep{Eisenstein2001}.  We choose  to apply our methodology to
UV-selected QSOs  and LRGs selected in the  2SLAQ \citep[2dF-SDSS LRGs
  and QSO;][]  {Cannon2006, Ross2007}  survey to estimate  the QSO/LRG
cross-correlation function and then infer the bias of the QSOs in that
survey.  We  refer the reader to  Appendix B for details  on the 2SLAQ
survey.  In  short, we use 448  spectroscopically identified UV-bright
QSOs and  $\approx10\,000$ LRGs  photometrically selected by  2SLAQ at
redshifts $0.35-0.75$.  We associate  the $\approx 10\,000$ 2SLAQ LRGs
with the  photometric redshift PDFs  estimated by Cunha et  al. (2009)
based on  the Sloan  Digital Sky Survey  (SDSS) DR7  photometry.  The
QSO/LRG real  space projected cross-correlation  function is estimated
using the LRG photometric redshift PDFs.  We then estimate the bias of
the tracer population using either  the full LRG sample or by applying
the  resampling   method  of  section   \ref{section:auto-cor}.   Both
approaches  give consistent  results for  the inferred  bias  of 2SLAQ
QSOs,   $b_Q=1.72^{+0.47}_{-0.44}$   and   $b_Q=1.58^{+0.50}_{-0.45}$,
respectively  (see   Appendix  B   for  details).  For  tracer
  populations with  relatively narrow redshift  distributions there is
  no need to follow the resampling approach to infer the clustering of
  the target population (i.e. QSOs).

\subsection{The clustering properties of X-ray AGN}

The     AGN/galaxy     cross-correlation     method     of     section
\ref{section:method}  is applied  to  the combined  AEGIS, COSMOS  and
ECDFS  fields.   The inferred  bias  of  X-ray  AGN is  compared  with
previous measurements in  Figure $\ref{fig:dmh}$.  The literature bias
datapoints in that figure  are presented in Table \ref{table:studies}.
They  are  re-estimated as  explained  in  Mountrichas \&  Georgakakis
(2012).  Our bias determinations  are consistent with previous studies
at similar $L_X$  and $z$ intervals that control  the impact of sample
variance either by  combining data from different surveys  or by using
the  AGN/galaxy cross-correlation function.   Indeed some  the highest
bias values of X-ray AGN  at $z\approx1$ in Figure $\ref{fig:dmh}$ are
because of serendipitous cosmic structures within the surveyed area
\cite[e.g.][]{Gilli2005, Gilli2009}.

The picture that emerges  from Figure $\ref{fig:dmh}$ is that moderate
luminosity  X-ray  AGN  reside  in   DMHs  with  mean  mass  of  about
$10^{13}h^{-1}$M$_\odot$ from $z\approx0$  to $z=1$. However, we
  also find evidence  that this average is skewed to  high values as a
  result of the distribution of AGN in halo mass. Excluding the 5 per
cent of  the X-ray  AGN in the  sample associated with  X-ray selected
groups, reduces the  inferred DMH mass of the  remaining population by
$\approx 60\%$ to $\approx 4 \times 10^{12}h^{-1}\,\rm M_{\odot}$.  It
is  emphasised that this  is not  because of  sample variance,  in the
sense  that   overdensities  within  the  surveyed   area  affect  the
clustering estimation. It is  rather a consequence of the way AGN
  are distributed in DMH  mass.  Our analysis therefore suggests that
a large fraction  of the moderate luminosity X-ray  AGN at $z\approx1$
live in DMHs with masses similar to those of UV bright QSOs, few times
$10^{12}h^{-1}\,\rm M_{\odot}$ \citep[e.g.][]{Croom2005, daAngela2008,
  Ross2009}.  This  DMH mass scale is consistent  with the predictions
of models which  invoke cold gas accretion via  either mergers or disk
instabilities to build SMBHs  at the centres of galaxies \citep[e.g][]
{Hopkins2007, Bonoli2009, Fanidakis2012}.   In contrast, if SMBHs grow
their mass  via accretion of  gas from a host  quasi-static atmosphere
\citep[e.g.]{Croton2006}   much   larger   DMH   mass   are   expected
\citep{Fanidakis2012}.   Our  analysis   therefore  favours  cold  gas
accretion as the main channel of SMBH growth for X-ray AGN.

\section{Conclusions}

A  method is  presented  to  estimate the  bias  of any  extragalactic
population, for which spectroscopy is available, via the estimation of
their real space cross-correlation  function with a tracer population,
for which only photometric redshift  PDFs are available. We argue that
this  method  works best  when  the  tracer  population has  a  narrow
redshift distribution that extends over  the same range as the sources
for which the clustering needs to be estimated.

The  method is applied  to moderate  luminosity [$L_X  (\rm 2-10\,keV)
\approx 10^{43} \,  erg \, s^{-1}$] X-ray AGN  at $z\approx1$ selected
in  AEGIS, COSMOS  and  ECDFS.  The  tracer  population are  optically
selected  galaxies in  those  fields.  DMH  masses  of $\rm  10^{13}\,
\,h^{-1}\,M_{\odot}$ are estimated for  the X-ray AGN samples, in good
agreement  with previous  studies at  similar redshift  and luminosity
intervals.

We also  find evidence that the distibution of AGN in dark matter halo
mass in skewed. After  excluding  5\%  of the  AGN  in the  sample
associated with X-ray  groups, we estimate a 0.5\,dex  lower DMH mass,
$\log M  \approx 12.5$  ($h^{-1}\,M_{\odot}$). Therefore, the  bulk of
the X-ray AGN live in  environments similar to those predicted by cold
gas  accretion models  for the  growth  of SMBHs,  e.g. gaseous  major
galaxy mergers or disk instabilities.

\section{Acknowledgments}

The authors are grateful to the anonymous referee for useful comments
and suggestions and Carie Cardamone for  providing photometric redshift
PDFs  for the  galaxies in  the MUSYC  survey field.   GM acknowledges
financial   support   from   the   Marie-Curie   Reintegration   Grant
PERG03-GA-2008-230644.   This study makes  use of  data from  AEGIS, a
multiwavelength sky survey conducted  with the Chandra, GALEX, Hubble,
Keck, CFHT,  MMT, Subaru, Palomar, Spitzer, VLA,  and other telescopes
and supported in part by the  NSF, NASA, and the STFC. Funding for the
DEEP3  Galaxy  Redshift  Survey   has  been  provided  by  NSF  grants
AST-0808133,  AST-0807630,  and  AST-0806732.  Based  on  observations
obtained  with   MegaPrime/MegaCam,  a  joint  project   of  CFHT  and
CEA/DAPNIA,  at the  Canada-France- Hawaii  Telescope (CFHT)  which is
operated  by  the  National  Research  Council (NRC)  of  Canada,  the
Institut National des Sciences de  l’Univers of the Centre National de
la  Recherche Scientifique  (CNRS) of  France, and  the  University of
Hawaii.   This work  is based  in part  on data  products  produced at
TERAPIX  and  the  Canadian  Astronomy  Data Centre  as  part  of  the
Canada-France- Hawaii Telescope Legacy Survey, a collaborative project
of NRC and CNRS.

\bibliography{mybib}{}

\begin{thebibliography}{64}
\expandafter\ifx\csname natexlab\endcsname\relax\def\natexlab#1{#1}\fi

\bibitem[{{Abazajian} {et~al.}(2009)}]{Abazajian2009}
{Abazajian} K.~N., {et~al.}, 2009, ApJS, 182, 543

\bibitem[{{Aird} {et~al.}(2010)}]{Aird2010}
{Aird} J., {et~al.}, 2010, MNRAS, 401, 2531

\bibitem[{{Allevato} {et~al.}(2011)}]{Allevato2011}
{Allevato} V., {et~al.}, 2011, ApJ, 736, 99

\bibitem[{{Allevato} {et~al.}(2012)}]{Allevato2012}
---, 2012, ApJ, 758, 47

\bibitem[{{Arnouts} {et~al.}(1999){Arnouts}, {Cristiani}, {Moscardini},
  {Matarrese}, {Lucchin}, {Fontana}, \& {Giallongo}}]{Arnouts1999}
{Arnouts} S., {Cristiani} S., {Moscardini} L., {Matarrese} S., {Lucchin} F.,
  {Fontana} A., {Giallongo} E., 1999, MNRAS, 310, 540

\bibitem[{{Bonoli} {et~al.}(2009){Bonoli}, {Marulli}, {Springel}, {White},
  {Branchini}, \& {Moscardini}}]{Bonoli2009}
{Bonoli} S., {Marulli} F., {Springel} V., {White} S.~D.~M., {Branchini} E.,
  {Moscardini} L., 2009, MNRAS, 396, 423

\bibitem[{{Brusa} {et~al.}(2010)}]{Brusa2010}
{Brusa} M., {et~al.}, 2010, ApJ, 716, 348

\bibitem[{{Cannon} {et~al.}(2006)}]{Cannon2006}
{Cannon} R., {et~al.}, 2006, MNRAS, 372, 425

\bibitem[{{Cappelluti} {et~al.}(2012){Cappelluti}, {Allevato}, \&
  {Finoguenov}}]{Cappelluti2012}
{Cappelluti} N., {Allevato} V., {Finoguenov} A., 2012, ArXiv e-prints,
  1201.3920

\bibitem[{{Cappelluti} {et~al.}(2009)}]{Cappelluti2009}
{Cappelluti} N., {et~al.}, 2009, A\&A, 497, 635

\bibitem[{{Cardamone} {et~al.}(2010)}]{Cardamone2010}
{Cardamone} C.~N., {et~al.}, 2010, ApJS, 189, 270

\bibitem[{{Coil} {et~al.}(2004)}]{Coil2004}
{Coil} A.~L., {et~al.}, 2004, ApJ, 609, 525

\bibitem[{{Coil} {et~al.}(2009)}]{Coil2009}
---, 2009, ApJ, 701, 1484

\bibitem[{{Cooper} {et~al.}(2011{\natexlab{a}})}]{Cooper2011_aces}
{Cooper} M.~C., {et~al.}, 2011{\natexlab{a}}, ArXiv e-prints, 1112.0312

\bibitem[{{Cooper} {et~al.}(2011{\natexlab{b}})}]{Cooper2011_deep3_1}
---, 2011{\natexlab{b}}, ApJS, 193, 14

\bibitem[{{Cooper} {et~al.}(2012)}]{Cooper2012_deep3_2}
---, 2012, MNRAS, 419, 3018

\bibitem[{{Coupon} {et~al.}(2009)}]{Coupon2009}
{Coupon} J., {et~al.}, 2009, A\&A, 500, 981

\bibitem[{{Croom} {et~al.}(2005)}]{Croom2005}
{Croom} S.~M., {et~al.}, 2005, MNRAS, 356, 415

\bibitem[{{Croton} {et~al.}(2006)}]{Croton2006}
{Croton} D.~J., {et~al.}, 2006, MNRAS, 365, 11

\bibitem[{{Cunha} {et~al.}(2009){Cunha}, {Lima}, {Oyaizu}, {Frieman}, \&
  {Lin}}]{Cunha2009}
{Cunha} C.~E., {Lima} M., {Oyaizu} H., {Frieman} J., {Lin} H., 2009, MNRAS,
  396, 2379

\bibitem[{{da \^{A}ngela} {et~al.}(2008)}]{daAngela2008}
{da \^{A}ngela} J., {et~al.}, 2008, MNRAS, 383, 565

\bibitem[{{Davis} {et~al.}(2007)}]{Davis2007}
{Davis} M., {et~al.}, 2007, ApJ, 660, L1

\bibitem[{{Eisenstein} {et~al.}(2001)}]{Eisenstein2001}
{Eisenstein} D.~J., {et~al.}, 2001, AJ, 122, 2267

\bibitem[{{Elvis} {et~al.}(2009)}]{Elvis2009}
{Elvis} M., {et~al.}, 2009, ApJS, 184, 158

\bibitem[{{Fanidakis} {et~al.}(2012)}]{Fanidakis2012}
{Fanidakis} N., {et~al.}, 2012, MNRAS, 419, 2797

\bibitem[{{Finoguenov} {et~al.}(2007)}]{Finoguenov2007}
{Finoguenov} A., {et~al.}, 2007, ApJS, 172, 182

\bibitem[{{Georgakakis} {et~al.}(2008){Georgakakis}, {Gerke}, {Nandra},
  {Laird}, {Coil}, {Cooper}, \& {Newman}}]{Georgakakis2008}
{Georgakakis} A., {Gerke} B.~F., {Nandra} K., {Laird} E.~S., {Coil} A.~L.,
  {Cooper} M.~C., {Newman} J.~A., 2008, MNRAS, 391, 183

\bibitem[{{Georgakakis} {et~al.}(2009)}]{Georgakakis2009}
{Georgakakis} A., {et~al.}, 2009, MNRAS, 397, 623

\bibitem[{{Georgakakis} {et~al.}(2011)}]{Georgakakis2011}
---, 2011, MNRAS, 418, 2590

\bibitem[{{George} {et~al.}(2011)}]{George2011}
{George} M.~R., {et~al.}, 2011, ApJ, 742, 125

\bibitem[{{Gilli} {et~al.}(2005)}]{Gilli2005}
{Gilli} R., {et~al.}, 2005, A\&A, 430, 811

\bibitem[{{Gilli} {et~al.}(2009)}]{Gilli2009}
---, 2009, A\&A, 494, 33

\bibitem[{{Groth} \& {Peebles}(1977)}]{Groth1977}
{Groth} E.~J., {Peebles} P.~J.~E., 1977, ApJ, 217, 385

\bibitem[{{Hickox} {et~al.}(2011)}]{Hickox2011}
{Hickox} R.~C., {et~al.}, 2011, ApJ, 731, 117

\bibitem[{{Hopkins} {et~al.}(2008){Hopkins}, {Hernquist}, {Cox}, \&
  {Keres}}]{Hopkins2008a}
{Hopkins} P.~F., {Hernquist} L., {Cox} T.~J., {Keres} D., 2008, ApJS, 175, 356

\bibitem[{{Hopkins} {et~al.}(2007){Hopkins}, {Lidz}, {Hernquist}, {Coil},
  {Myers}, {Cox}, \& {Spergel}}]{Hopkins2007}
{Hopkins} P.~F., {Lidz} A., {Hernquist} L., {Coil} A.~L., {Myers} A.~D., {Cox}
  T.~J., {Spergel} D.~N., 2007, ApJ, 662, 110

\bibitem[{{Ilbert} {et~al.}(2006)}]{Ilbert2006}
{Ilbert} O., {et~al.}, 2006, A\&A, 457, 841

\bibitem[{{Krumpe} {et~al.}(2010){Krumpe}, {Miyaji}, \& {Coil}}]{Krumpe2010}
{Krumpe} M., {Miyaji} T., {Coil} A.~L., 2010, ApJ, 713, 558

\bibitem[{{Krumpe} {et~al.}(2012){Krumpe}, {Miyaji}, {Coil}, \&
  {Aceves}}]{Krumpe2012}
{Krumpe} M., {Miyaji} T., {Coil} A.~L., {Aceves} H., 2012, ApJ, 746, 1

\bibitem[{{Laird} {et~al.}(2009)}]{Laird2009}
{Laird} E.~S., {et~al.}, 2009, ApJS, 180, 102

\bibitem[{{Landy} \& {Szalay}(1993)}]{LZ1993}
{Landy} S.~D., {Szalay} A.~S., 1993, ApJ, 412, 64

\bibitem[{{Leauthaud} {et~al.}(2010)}]{Leauthaud2010}
{Leauthaud} A., {et~al.}, 2010, ApJ, 709, 97

\bibitem[{{Lehmer} {et~al.}(2005)}]{Lehmer2005}
{Lehmer} B.~D., {et~al.}, 2005, ApJS, 161, 21

\bibitem[{{Lilly} {et~al.}(2009)}]{Lilly2009}
{Lilly} S.~J., {et~al.}, 2009, ApJS, 184, 218

\bibitem[{{Limber}(1953)}]{Limber1953}
{Limber} D.~N., 1953, ApJ, 117, 134

\bibitem[{{Marulli} {et~al.}(2008){Marulli}, {Bonoli}, {Branchini},
  {Moscardini}, \& {Springel}}]{Marulli2008}
{Marulli} F., {Bonoli} S., {Branchini} E., {Moscardini} L., {Springel} V.,
  2008, MNRAS, 385, 1846

\bibitem[{{Miyaji} {et~al.}(2011){Miyaji}, {Krumpe}, {Coil}, \&
  {Aceves}}]{Miyaji2011}
{Miyaji} T., {Krumpe} M., {Coil} A.~L., {Aceves} H., 2011, ApJ, 726, 83

\bibitem[{{Miyaji} {et~al.}(2007)}]{Miyaji2007}
{Miyaji} T., {et~al.}, 2007, ApJS, 172, 396

\bibitem[{{Mo} \& {White}(1996)}]{Mo1996}
{Mo} H.~J., {White} S.~D.~M., 1996, MNRAS, 282, 347

\bibitem[{{Mountrichas} \& {Georgakakis}(2012)}]{Mountrichas2012}
{Mountrichas} G., {Georgakakis} A., 2012, MNRAS, 420, 514

\bibitem[{{Mountrichas} {et~al.}(2009){Mountrichas}, {Sawangwit}, {Shanks},
  {Croom}, P., {Myers}, \& {Pimbblet}}]{Mountrichas2009}
{Mountrichas} G., {Sawangwit} U., {Shanks} T., {Croom} S.~M., P. S.~D., {Myers}
  A.~D., {Pimbblet} K., 2009, MNRAS, 394, 2050

\bibitem[{{Myers} {et~al.}(2009){Myers}, {White}, \& {Ball}}]{Myers2009}
{Myers} A.~D., {White} M., {Ball} N.~M., 2009, MNRAS, 399, 2279

\bibitem[{{Newman} {et~al.}(2012)}]{Newman2012}
{Newman} J.~A., {et~al.}, 2012, ArXiv e-prints, 1203.3192

\bibitem[{{Pickles}(1998)}]{Pickles1998}
{Pickles} A.~J., 1998, PASP, 110, 863

\bibitem[{{Richards} {et~al.}(2005)}]{Richards2005}
{Richards} G.~T., {et~al.}, 2005, MNRAS, 360, 839

\bibitem[{{Ross} {et~al.}(2011){Ross}, {Tojeiro}, \& {Percival}}]{RossA2011}
{Ross} A.~J., {Tojeiro} R., {Percival} W.~J., 2011, MNRAS, 413, 2078

\bibitem[{{Ross} {et~al.}(2007)}]{Ross2007}
{Ross} N.~P., {et~al.}, 2007, MNRAS, 381, 573

\bibitem[{{Ross} {et~al.}(2009)}]{Ross2009}
---, 2009, ApJ, 697, 1634

\bibitem[{{Scoville} {et~al.}(2007)}]{Scoville2007}
{Scoville} N., {et~al.}, 2007, ApJS, 172, 1

\bibitem[{{Sheth} {et~al.}(2001){Sheth}, {Mo}, \& {Tormen}}]{Sheth2001}
{Sheth} R.~K., {Mo} H.~J., {Tormen} G., 2001, MNRAS, 323, 1

\bibitem[{{Silverman} {et~al.}(2010)}]{Silverman2010}
{Silverman} J.~D., {et~al.}, 2010, ApJS, 191, 124

\bibitem[{{Sutherland} \& {Saunders}(1992)}]{Sutherland_and_Saunders1992}
{Sutherland} W., {Saunders} W., 1992, MNRAS, 259, 413

\bibitem[{{Trump} {et~al.}(2009)}]{Trump2009}
{Trump} J.~R., {et~al.}, 2009, ApJ, 696, 1195

\bibitem[{{Van den Bosch}(2002)}]{Bosch2002}
{Van den Bosch} F.~C., 2002, MNRAS, 331, 98

\end{thebibliography}
\bibliographystyle{mn2e}

\appendix

\chapter{First Appendix}

\section{Colour transformations to the DEEP2 BRI filterset}

The analysis  presented in  this paper exploits  the fact  that simple
colour cuts  provide an  efficient way of  selecting galaxies  in well
defined redshift  intervals.  We choose  to study AGN in  the redshift
range  $0.7-1.4$  and therefore  adopt  the  DEEP2 survey  photometric
criteria for  redshift pre-selection \citep{Newman2012}.   They showed
that galaxies  below and above $z=0.7$  separate well in  the $B-R$ vs
$R-I$  colour  space,  thereby  yielding nearly  complete  samples  of
galaxies at $z>0.7$.

Applying the  DEEP2 colour  cuts to the  MUSYC and  CFHTLS photometric
catalogues  requires the determination  of transformations  between the
filtersets of those surveys and the DEEP2 $BRI$ photometric bands.

\subsection{The MUSYC dataset}

\begin{figure*}
\begin{center}
\includegraphics[height=0.9\columnwidth]{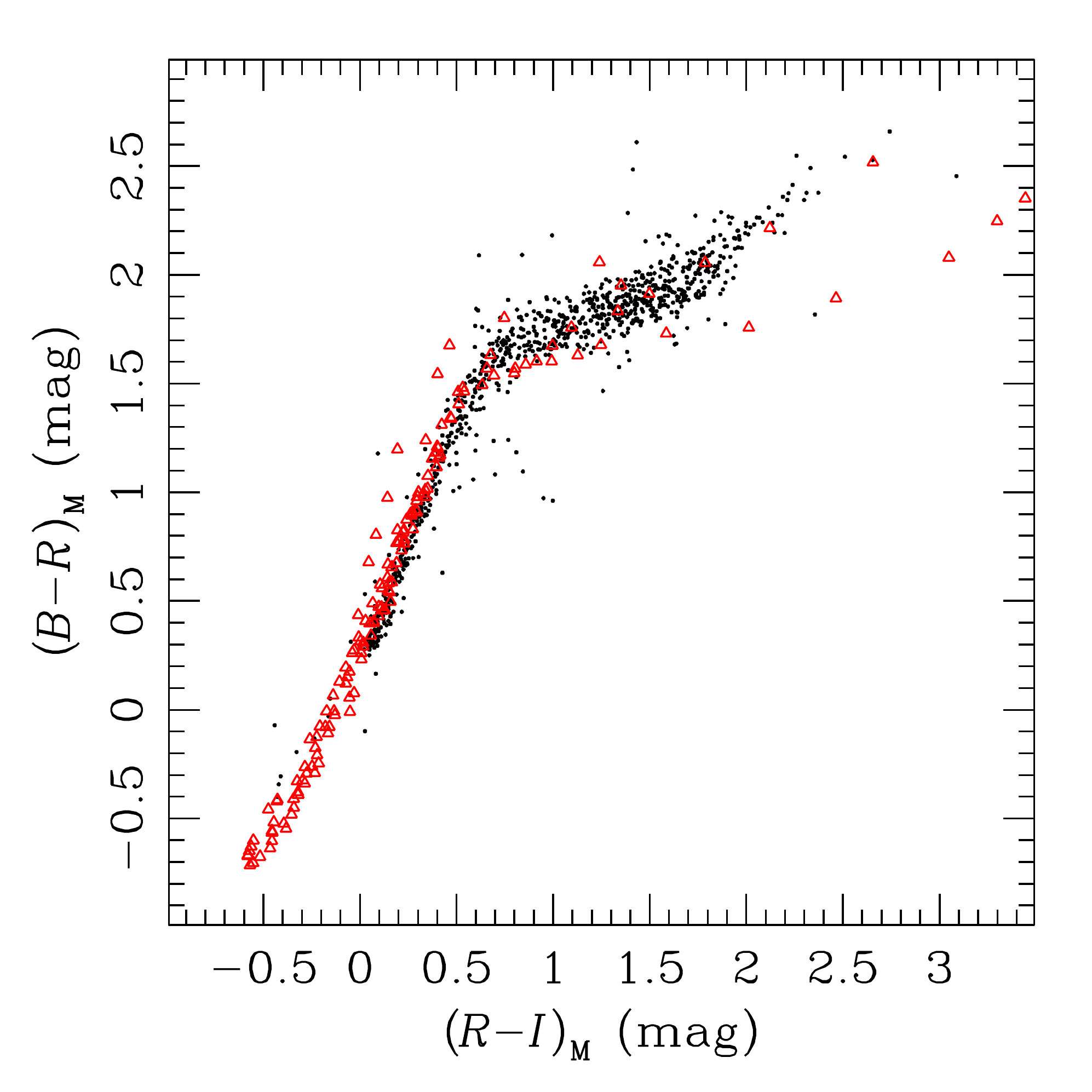}
\includegraphics[height=0.9\columnwidth]{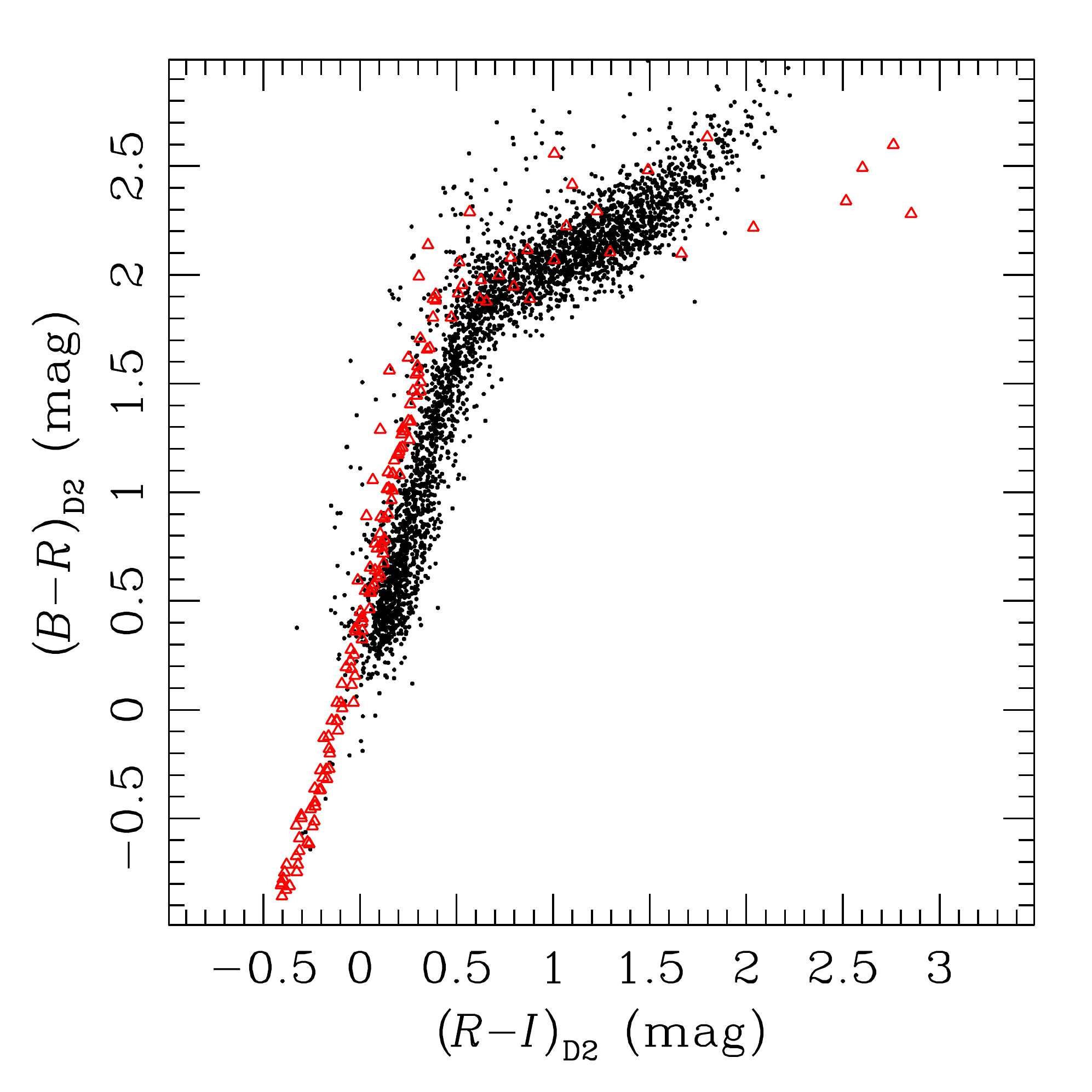}
\end{center}
\caption{$B-R$ vs  $R-I$ stellar locus. The  panel on the  left is for
  the MUSYC  filterset (subscript ``M''  in $B-R$ and  $R-I$ colours).
  The right panel is for  the DEEP2 bands (subscript ``D2'').  In both
  panels the  black dots are optically unresolved  sources detected in
  the  MUSYC (left) and  DEEP2 (right)  photometric surveys.   The red
  triangles   are  the  theoretical   stellar  tracks   determined  by
  convolving  the Pickles  (1998)  stellar template  library with  the
  MUSYC (left panel) and DEEP2 (right panel) $BRI$ filters.}
\label{fig_musyc_deep2_stars}
\end{figure*}

In  the ECDFS  we use  the MUSYC  Subaru v1.0  catalog  which provides
photometry   in   32  bands   from   the   UV   to  the   mid-infrared
\citep{Cardamone2010}.  For the DEEP2 we use the photometric catalogue
of \cite{Coil2004}.  As there is  no overlap between the MUSYC and any
of the DEEP2 fields, the colour transformations between the filtersets
in  the  two  photometric  surveys are  determined  using  theoretical
stellar  tracks. The LePHARE  code \citep{Arnouts1999,  Ilbert2006} is
used to convolve the DEEP2 and MUSYC $B$, $R$ and $I$ filters with the
\cite{Pickles1998} stellar  templates and predict the  $B-R$ and $R-I$
colours    of   stars    in   the    two   filtersets.     In   Figure
\ref{fig_musyc_deep2_stars} these stellar tracks are compared with the
photometry of optically unresolved sources (i.e. stellar like) in each
of  the two  surveys.  In  the DEEP2  catalogue these  are  defined as
objects with  $P_{GAL}<0.2$ (probability that the  source is resolved,
see  \citealp  {Coil2004} for  details)  and $19<R_{\rm  D2}<23$\,mag,
where  $R_{\rm D2}$  is the  optical magnitude  measured in  the DEEP2
$R$-band  filter.   In the  case  of  the  MUSYC catalogue  stars  are
selected by requiring that the parameter {\sc star\_flag} equals unity
(see \citealp {Cardamone2010} for details) and $19<R_{\rm M}<23$\,mag,
where  $R_{\rm M}$  is the  optical  magnitude measured  in the  MUSYC
$R$-band  filter.  Figure  \ref{fig_musyc_deep2_stars} shows  that the
photometric calibration of the MUSYC  and DEEP2 surveys is offset from
the  theoretical  stellar tracks  by  about  $+0.05$ and  $+0.15$\,mag
respectively in the $R-I$  colour.  These systematic offsets should be
taken  into  account when  converting  from  the  MUSYC to  the  DEEP2
filterset.  The  source of  these offsets is  beyond the scope  of the
analysis presented here.

\begin{figure*}
\begin{center}
\includegraphics[height=0.9\columnwidth]{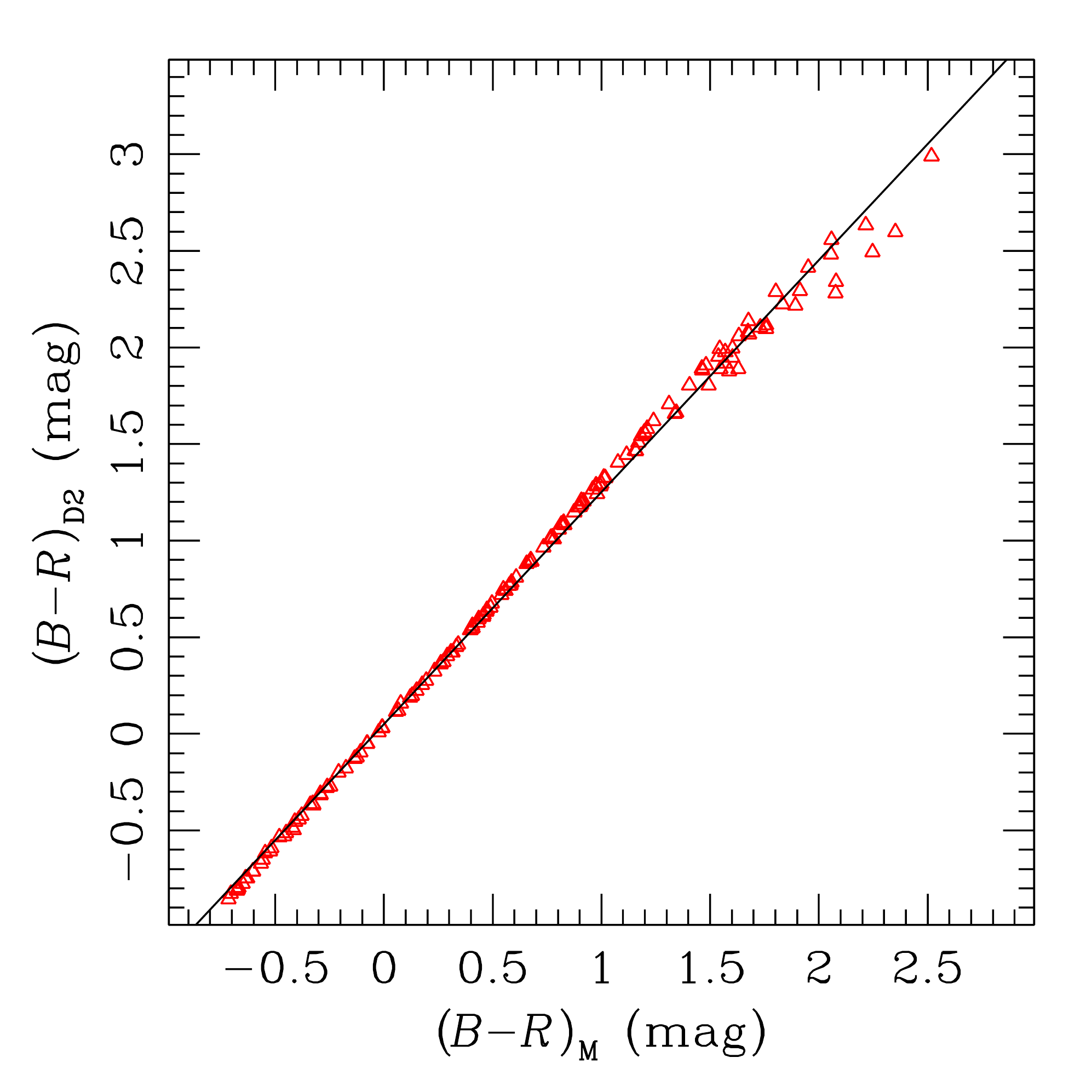}
\includegraphics[height=0.9\columnwidth]{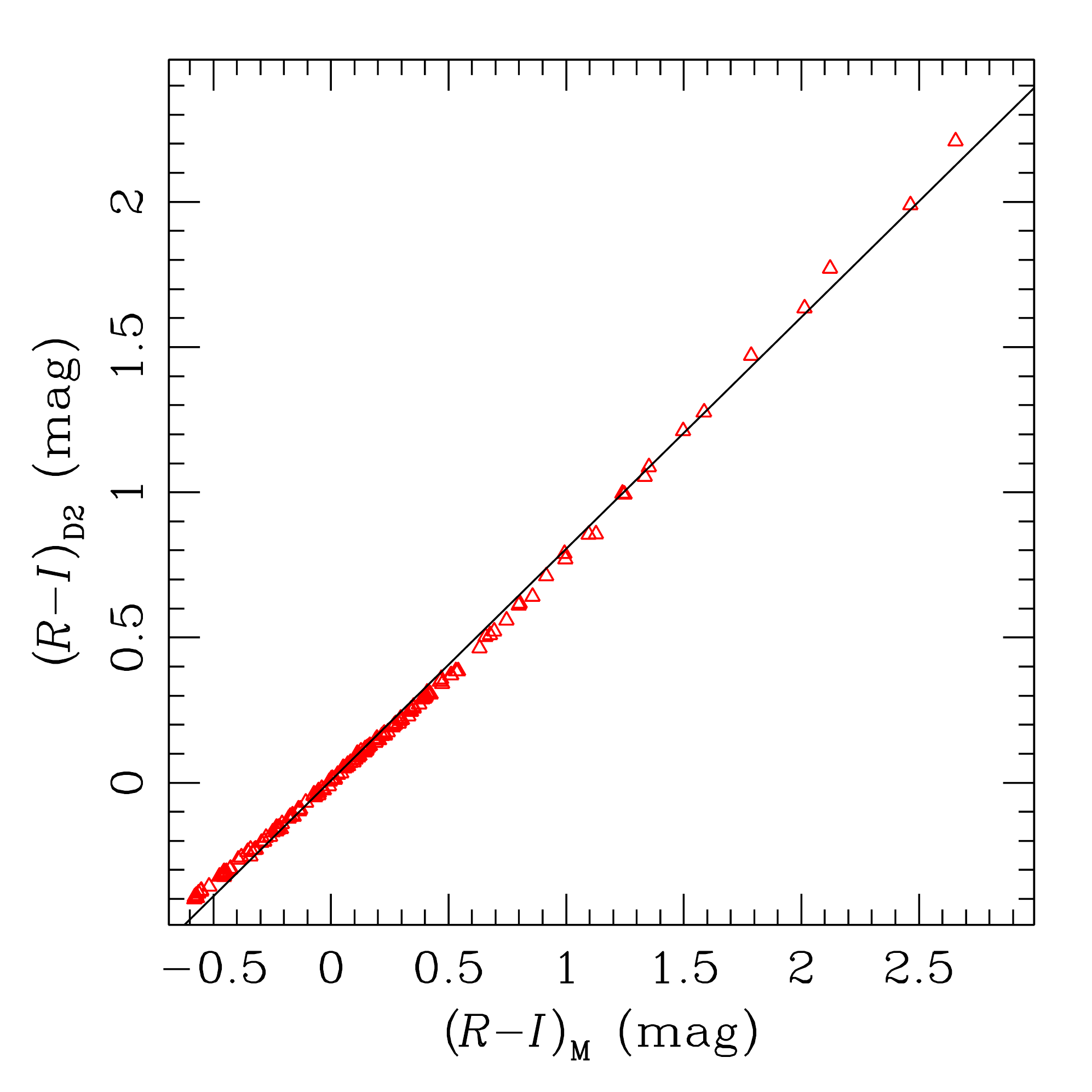}
\end{center}
\caption{{\bf  Left panel:} $(B-R)_{\rm  M}$ colour  estimated through
  the MUSYC  bands against $(B-R)_{\rm D2}$  colour determined through
  the DEEP2 filters.  The red triangles are the Pickles (1998) stellar
  templates. The black line shows  the best-fit linear relation to the
  theoretical stellar track. {\bf Right panel:} $(R-I)_{\rm M}$ colour
  estimated through  the MUSYC  bands against $(R-I)_{\rm  D2}$ colour
  determined  through the DEEP2  filters.  The  red triangles  are the
  Pickles (1998) stellar templates.  The black line shows the best-fit
  linear      relation       to      the      theoretical      stellar
  track.}\label{fig_musyc_deep2}
\end{figure*}

Next we compare  the MUSYC and DEEP2 filtersets  by plotting in Figure
\ref{fig_musyc_deep2} the $(B-R)_{\rm  M}$ versus the $(B-R)_{\rm D2}$
and the  $(R-I)_{\rm M}$ against  the $(R-I)_{\rm D2}$ colours  of the
\cite{Pickles1998}  stellar  templates.   In  the notation  above  the
subscripts  $\rm M$  and $\rm  D2$ denote  the MUSYC  and  DEEP2 bands
respectively.  For simplicity, linear relations are fit to the stellar
tracks  in Figure \ref{fig_musyc_deep2}  to transform  the $(B-R)_{\rm
M}$ and  $(R-I)_{\rm M}$ colours  to $(B-R)_{\rm D2}$  and $(R-I)_{\rm
D2}$  respectively.   Fitting higher  order  polynomials  to the  data
points in  Figure \ref{fig_musyc_deep2}  does not change  the results.
The best-fit linear relations are

\begin{equation}
\label{BRd2_to_BRm} (B-R)_{\rm D2}=0.05 + 1.2\,(B-R)_{\rm M},
\end{equation}

\begin{equation}
\label{RId2_to_RIm}     
(R-I)_{\rm     D2}=0.007     +
0.8\,(R-I)_{\rm M}.
\end{equation}

\begin{figure}
\begin{center}
\includegraphics[height=0.9\columnwidth]{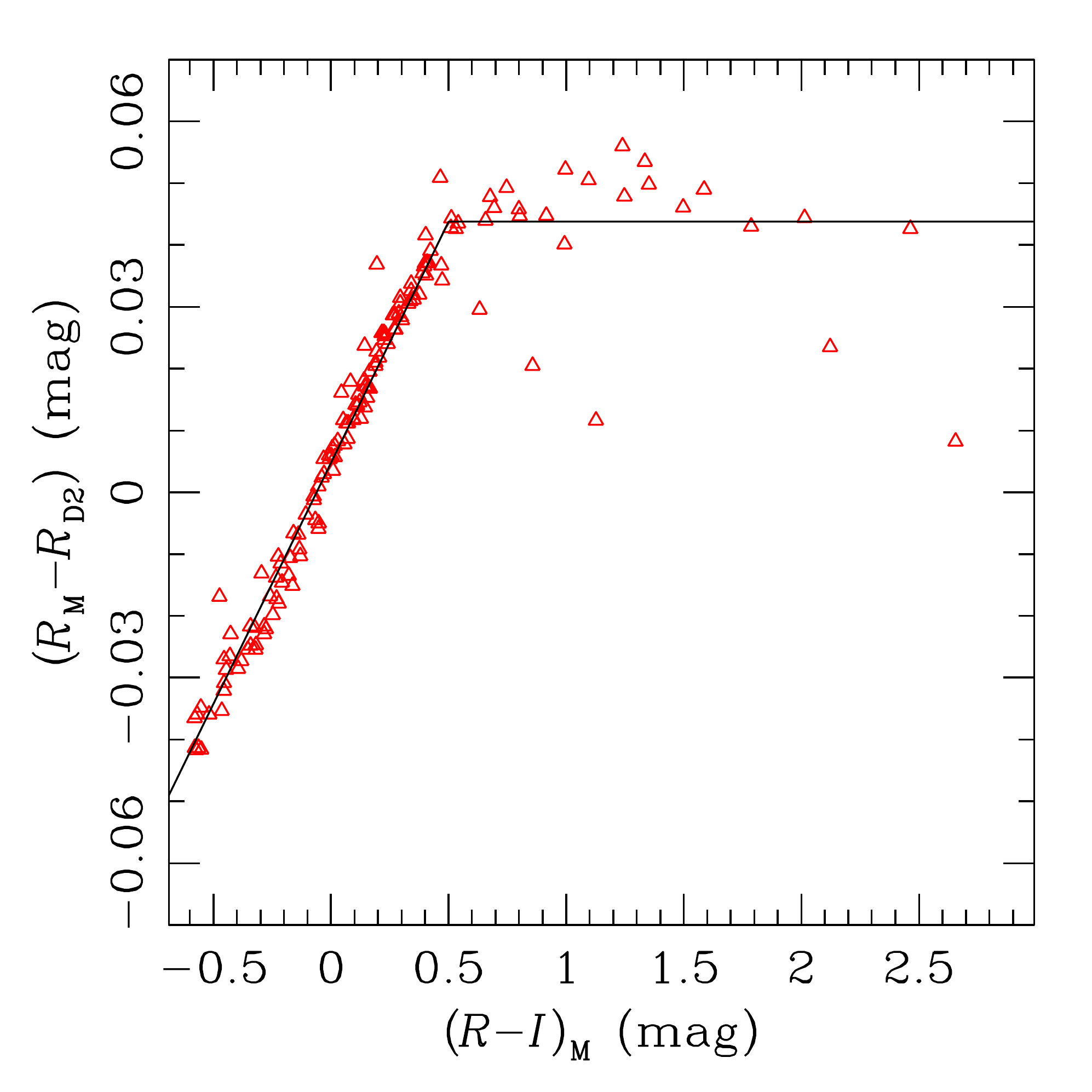}
\end{center}
\caption{$R_{\rm  M}-R_{\rm D2}$ against  $(R-I)_{\rm M}$  colour. The
  red triangles  are the Pickles  (1998) stellar templates.  The black
  line shows the relation  used to approximate the theoretical stellar
  track.  }\label{fig_musyc_deep2_rband}
\end{figure}

\noindent  Figure \ref{fig_musyc_deep2_rband} plots  $R_{\rm M}-R_{\rm
D2}$  against  $(R-I)_{\rm M}$  for  the  \cite {Pickles1998}  stellar
templates.  It  shows that $R_{\rm  M}-R_{\rm D2}$ is  well correlated
with the  $(R-I)_{\rm M}$ colour  up to $(R-I)_{\rm  M}\approx0.5$. At
redder colours the relation appears  to flatten and the scatter in the
theoretical  stellar  locus increases.   We  choose  to  fit a  linear
relation  to the  datapoints  up to  $(R-I)_{\rm  M}=0.5$.  At  redder
colours, the  $R_{\rm M}-R_{\rm D2}$  is approximated with  a constant
fixed  to the  value of  the linear  relation at  $(R-I)_{\rm M}=0.5$.
Although this  approximation may  lead to systematic  uncertainties in
the conversion from $R_{\rm M}$ to $R_{\rm D2}$, these are expected to
be at  the level of  few tenths of  a magnitude.  The  functional form
used to  describe the  variations of the  $R_{\rm M}-R_{\rm  D2}$ with
$(R-I)_{\rm M}$ is

\begin{equation}\label{Rd2_to_Rm}
R_{\rm  D2}-R_M =  \left\{\begin{array}{ll} 0.078  + 0.005\,(R-I)_{\rm
  M}, & (R-I)_{\rm M}<0.5, \\
\\ 0.04, & (R-I)_{\rm M}>0.5.\\
\end{array} \right.
\end{equation}

\begin{figure}
\begin{center}
\includegraphics[height=0.9\columnwidth]{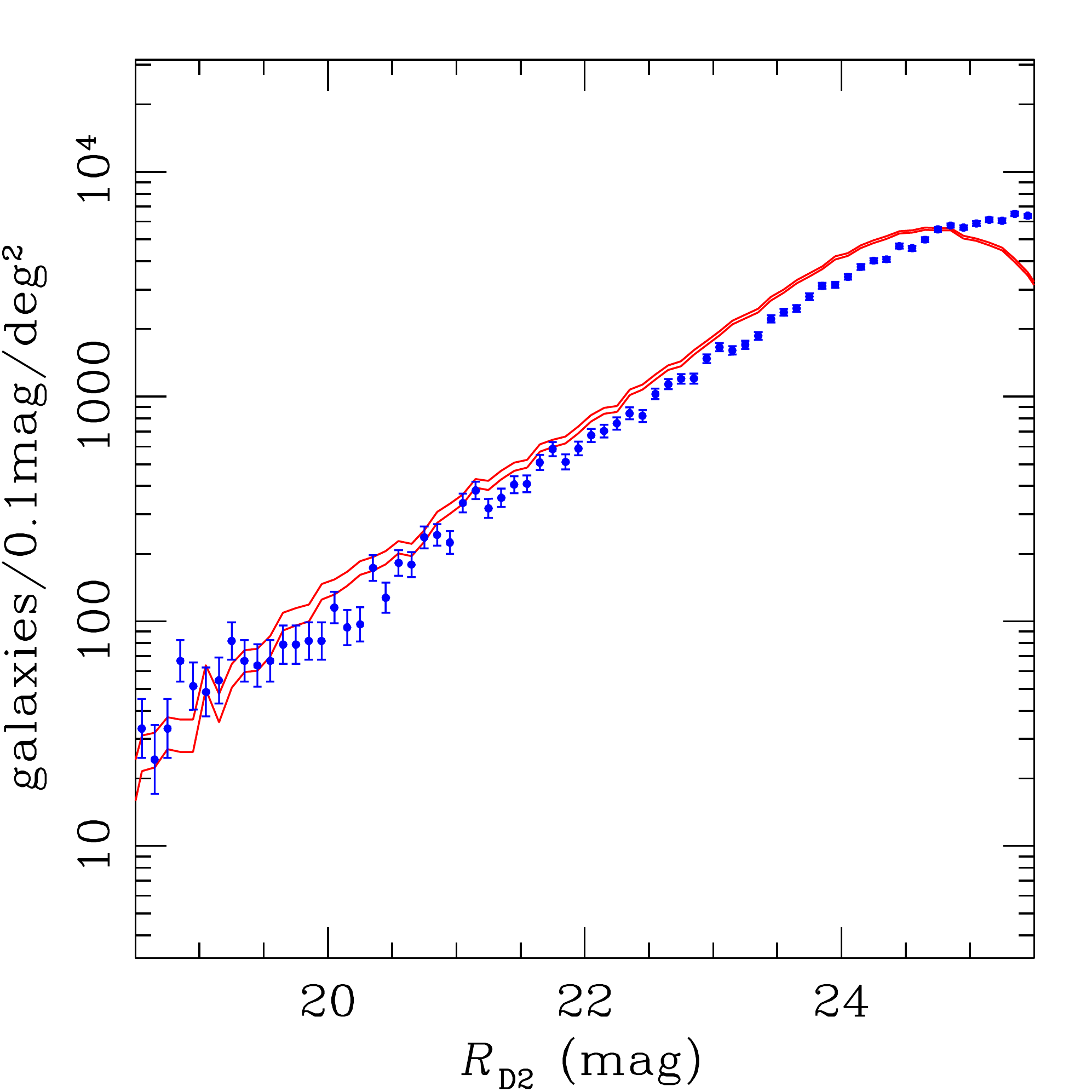}
\end{center}
\caption{Galaxy number  counts in 0.1\,mag  bins for the  $R_{\rm D2}$
  filter.  The red  curves correspond to the 1\,sigma  rms envelope of
  the galaxy  counts in the  DEEP2 photometric survey of  the Extended
  Groth  Strip.   These observations  become  incomplete fainter  than
  $R\approx24$\,mag  resulting  in  the  observed  turnover  at  faint
  magnitudes.  The blue dots  are for the MUSYC photometric catalogue,
  in  which  case  $R_M$  is  converted  to  $R_{D2}$  using  equation
  \ref{Rd2_to_Rm}.  The number counts  are normalised  to the  area of
  each  survey,  $\rm 0.33\,deg^2$  in  the  case  of MUSYC  and  $\rm
  1.18\,deg^2$ for the DEEP2 survey of the Extended Groth Strip. There
  is  systematic offset  of about  0.2\,mag between  the  two surveys.
}\label{fig_musyc_deep2_counts}
\end{figure}

\begin{figure}
\begin{center}
\includegraphics[height=0.9\columnwidth]{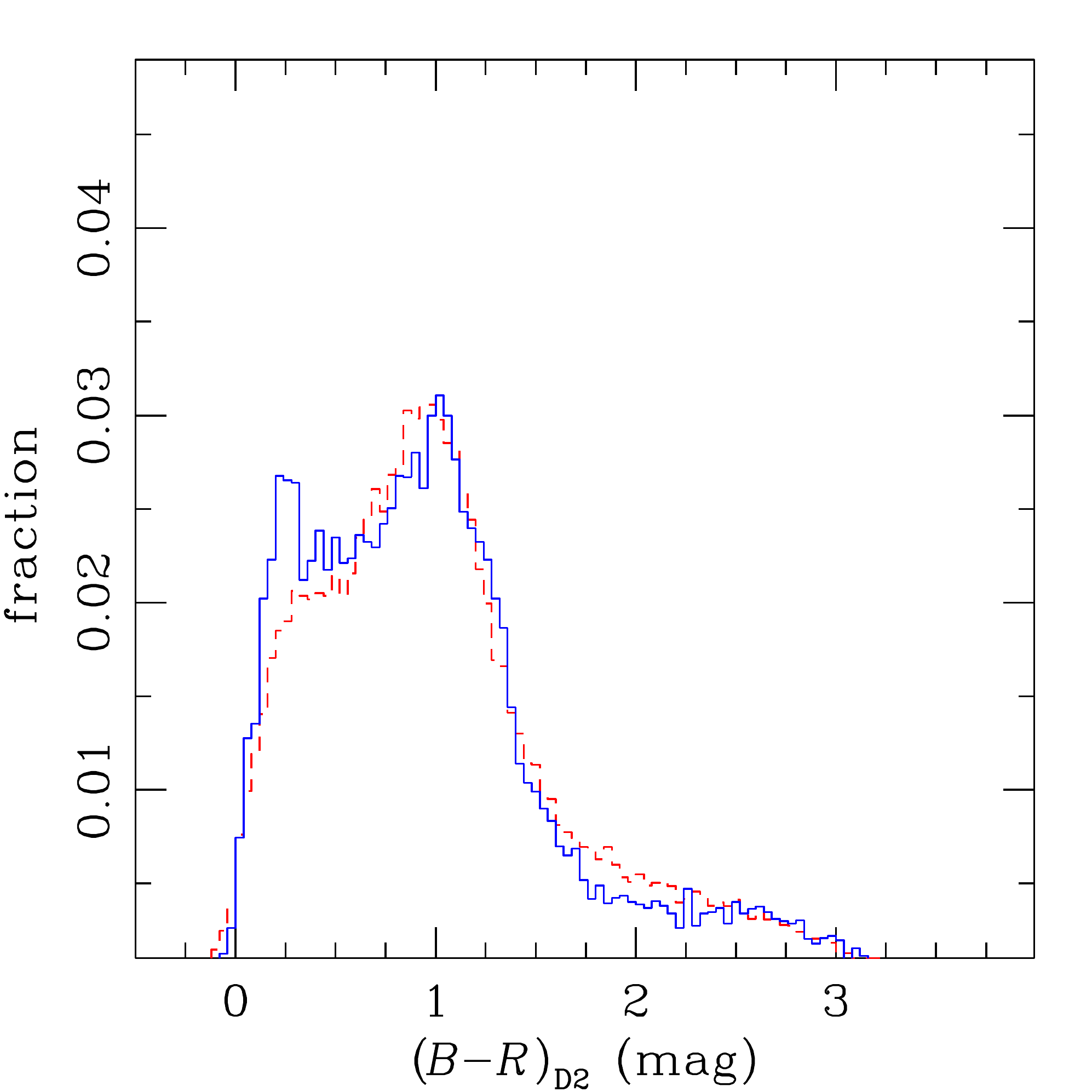}
\includegraphics[height=0.9\columnwidth]{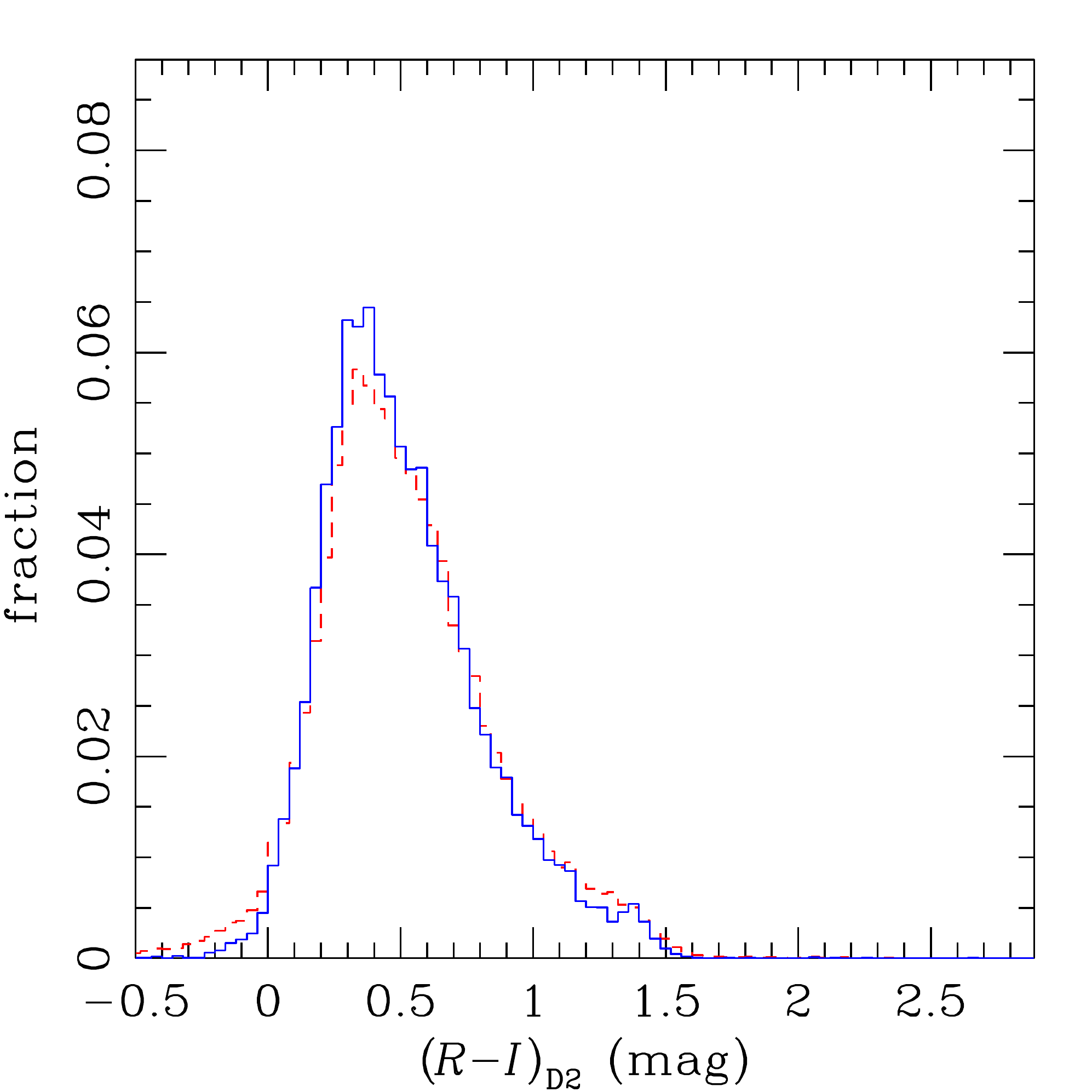}
\end{center}
\caption{$(B-R)_{\rm  D2}$ and  $(R-I)_{D2}$  colour distributions  in
  0.40\,mag bins of  galaxies selected in the MUSYC  (blue solid line)
  and  DEEP2 (red hatched  histogram) surveys.  For MUSYC  sources the
  colours   are    estimated   using   the    colour   transformations
  \ref{BRd2_to_BRm},  \ref{RId2_to_RIm}.  The  colour  distribution is
  limited to galaxies in the two surveys with $18<R_{D2}<24$\,mag.}
\label{fig_musyc_deep2_colours}
\end{figure}

\noindent  As  a  consistency  check  of  the  colour  transformations
\ref{BRd2_to_BRm},  \ref{RId2_to_RIm}, \ref{Rd2_to_Rm} we  compare the
magnitude and colour distributions of  galaxies in the DEEP2 and MUSYC
surveys.   For  each source  in  the  MUSYC  photometric catalogue  we
estimate  its  $(B-R)_{\rm  D2}$,  $(R-I)_{D2}$ colours  and  $R_{D2}$
magnitude   by  applying   the  linear   relations  \ref{BRd2_to_BRm},
\ref{RId2_to_RIm}, \ref{Rd2_to_Rm}.  The number counts in the $R_{D2}$
band  are then constructed  for MUSYC  sources classified  as galaxies
({\sc star\_flag} parameter  equals null; see \citealp {Cardamone2010}
for details).   These are then compared with  the $R_{D2}$-band number
counts  of  DEEP2  galaxies  (extended  source  probability  parameter
$P_{GAL}>0.2$; see Coil et al.   2004 for details).  The galaxy counts
normalised  to the  area of  each  photometric survey  are plotted  in
Figure \ref{fig_musyc_deep2_counts}.  There  is a systematic offset of
about  0.2\,mag  between  the  two  photometric  catalogues.   Similar
offsets are also apparent in  the $B_{D2}$ and $I_{D2}$ galaxy counts.
This suggests  that the discrepancy  is because of differences  in the
determination  of total  magnitudes  in  the two  surveys  and is  not
related to the colour  transformation between the two filtersets. This
is  further supported  by  Figure \ref{fig_musyc_deep2_colours}  which
compares the  $(B-R)_{\rm D2}$  and $(R-I)_{D2}$ distributions  of the
MUSYC and DEEP2 surveys.   These are constructed by selecting galaxies
in the two surveys with $18<R_{D2}<24$\,mag. The faint magnitude limit
is chosen  to minimise the  effect of incompleteness in  the shallower
DEEP2  photometric survey.   In the  case  of the  MUSYC catalogue  an
offset of  +0.2\,mag is applied  to the estimated $R_{D2}$  to account
for the systematic offset in the number counts of the two surveys.  In
the case  of the $(R-I)_{D2}$ colour distribution  the offsets between
the theoretical stellar tracks and the MUSYC/DEEP2 photometry (see
Fig.   \ref{fig_musyc_deep2_stars})  have  also  been applied  to  the
data. The  agreement between the MUSYC and  DEEP2 colour distributions
in  Figure  \ref{fig_musyc_deep2_colours}  indicates that  the  colour
transformations of  equations \ref{BRd2_to_BRm}, \ref{RId2_to_RIm} are
robust  and can be  used to  convert the  fluxes measured  through the
MUSYC filterset to the DEEP2 $BRI$ bands.

\subsection{The CFHTLS dataset}

The CFHTLS  D3 field  overlaps with the  DEEP2 survey in  the Extended
Groth  Strip  (EGS).   It  is  therefore  possible  to  determine  the
transformations  from  the CFHTLS  $ugriz$  bands  to  the DEEP2  $BRI$
filterset using sources that are  common in the two surveys.  The data
release T0004  of the CFHTLS  photometric catalogue \citep{Coupon2009}
is cross-matched  with the DEEP2 source list  \citep{Coil2004} using a
search  radius of  2\,arcsec. Regions  of poor  photometry  because of
nearby  bright stars  or diffraction  spikes  are masked  out in  this
exercise.

\begin{figure*}   
\begin{center}
\includegraphics[height=0.9\columnwidth]{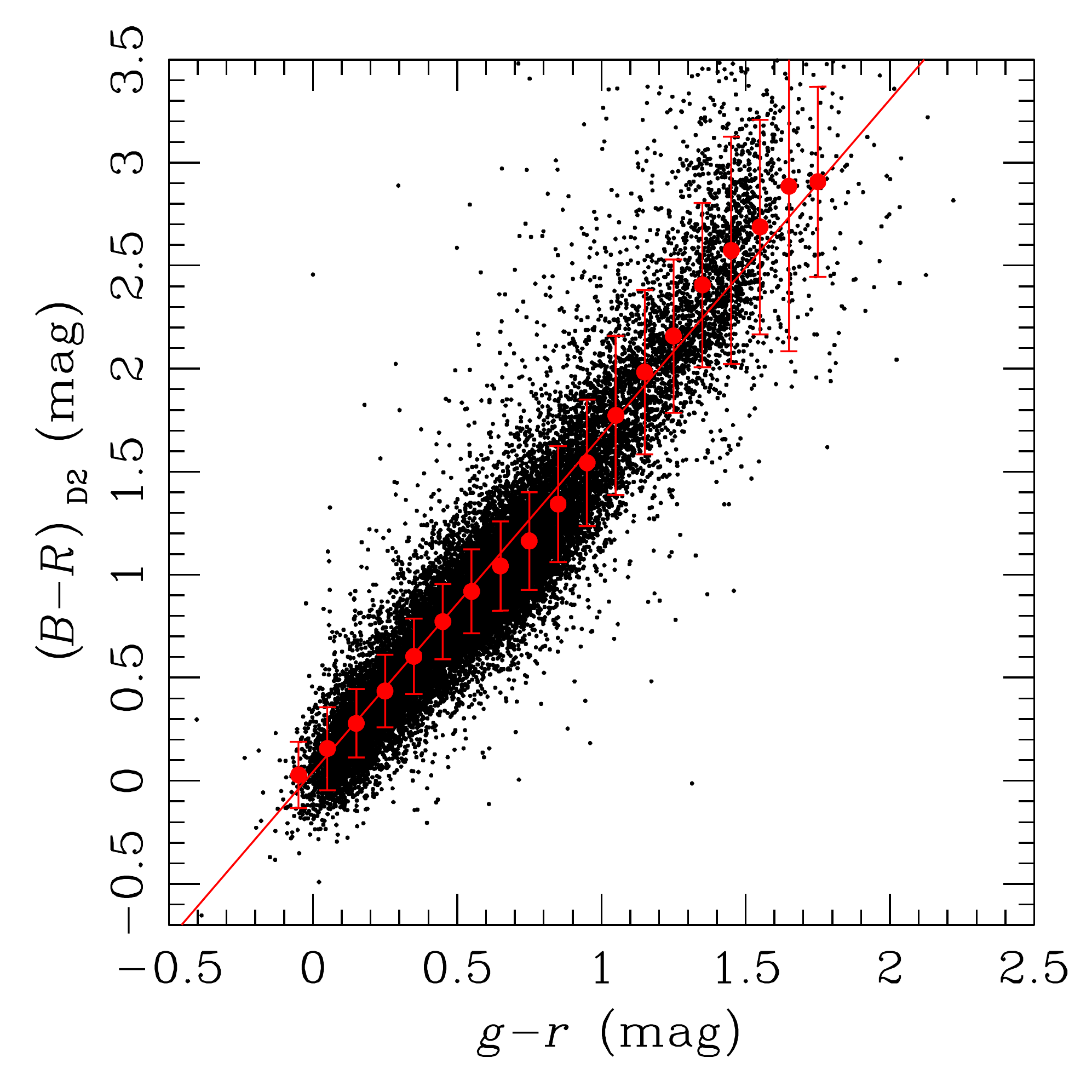}
\includegraphics[height=0.9\columnwidth]{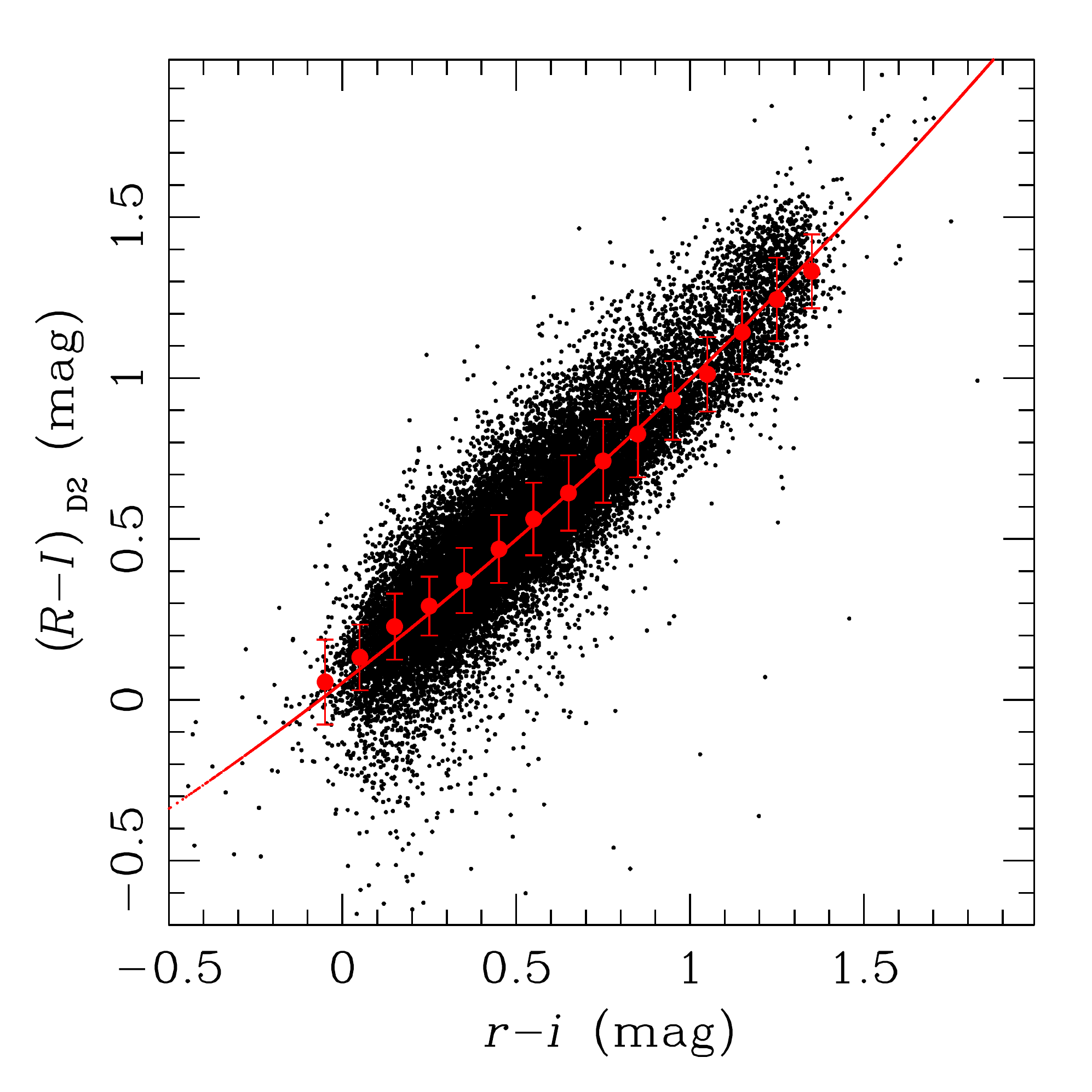}
\end{center}
\caption{{\bf   Left:}   $(B-R)_{\rm   D2}$  plotted   against   $g-r$
  colour. The black dots are resolved  sources in the D3 region of the
  CFHTLS  (parameter {\sc  flag\_terapix} equals  null, Coupon  et al.
  2009) with  DEEP2 photometric survey counterparts  and magnitudes in
  the  range  $18<R_{D2}<24$\,mag.   The  mean  $(B-R)_{\rm  D2}$  and
  1\,sigma rms  within $g-r$ bins  (0.1\,mag size) are shown  with the
  red circles  and errorbars.   The red line  is the  best-fit linear
  relation to  those datapoints. The  rms scatter around this  line is
  estimated  to be  about  0.15\,mag.  {\bf  Right:} $(R-I)_{\rm  D2}$
  plotted against $r-i$ colour. The black dots are resolved sources in
  the D3 region of the CFHTLS with DEEP2 counterparts in the magnitude
  range  $18<R_{D2}<24$\,mag. The mean  $(R-I)_{\rm D2}$  and 1\,sigma
  rms within $r-i$ bins (0.1\,mag size) are shown with the red circles
  and  errorbars.   The  red   curve  is  the  best-fit  second  order
  polynomial to those datapoints. The  rms scatter around this line is
  estimated to be about 0.1\,mag. }
\label{fig_cfhtls_deep2}
\end{figure*}

The transformations of the $g-r$ and $r-i$ colours to $(B-R)_{\rm D2}$
and $(R-I)_{\rm  D2}$  respectively  are determined  using  optically
resolved  sources  in  the  deeper CFHTLS  catalogue  (parameter  {\sc
  flag\_terapix}    equals   null,    \citealp    {Coupon2009})   with
$18<R_{D2}<24$\,mag.   We  choose to  use  galaxies  to determine  the
transformation between colours in the  two filtersets to have a handle
on  the scatter  expected in  those relations  because  of photometric
uncertainties, the different approaches for determining colours in the
two  surveys  as  well  as   the  diversity  of  the  Spectral  Energy
Distributions of extragalactic sources.  Figure \ref{fig_cfhtls_deep2}
compares the CFHTLS  and DEEP2 filtersets by plotting  for galaxies in
the  overlap region  of the  two surveys  the $(B-R)_{\rm  D2}$ colour
versus $g-r$ and the $(R-I)_{\rm  D2}$ colour against $r-i$.  The mean
$(B-R)_{\rm D2}$, [$(R-I)_{\rm D2}$] and its standard deviation within
$g-r$ ($r-i$)  colour bins of 0.1\,mag  size are also  shown in Figure
\ref{fig_cfhtls_deep2}.   A  second order  polynomial  is  fit to  the
binned datapoints in  the $(R-I)_{\rm D2}$ versus $r-i$  plot, while a
linear relation is used for  the $(B-R)_{\rm D2}$ against $g-r$ colour
plot.

\begin{equation}\label{BRd2_to_gr}
(B-R)_{\rm D2}=0.04 + 1.63\,(g-r),
\end{equation}

\begin{equation}\label{RId2_to_ri}
(R-I)_{\rm D2}=0.054+ 0.84 \,(r-i) + 0.11 \, (r-i)^2.
\end{equation}

\begin{figure}
\begin{center}
\includegraphics[height=0.9\columnwidth]{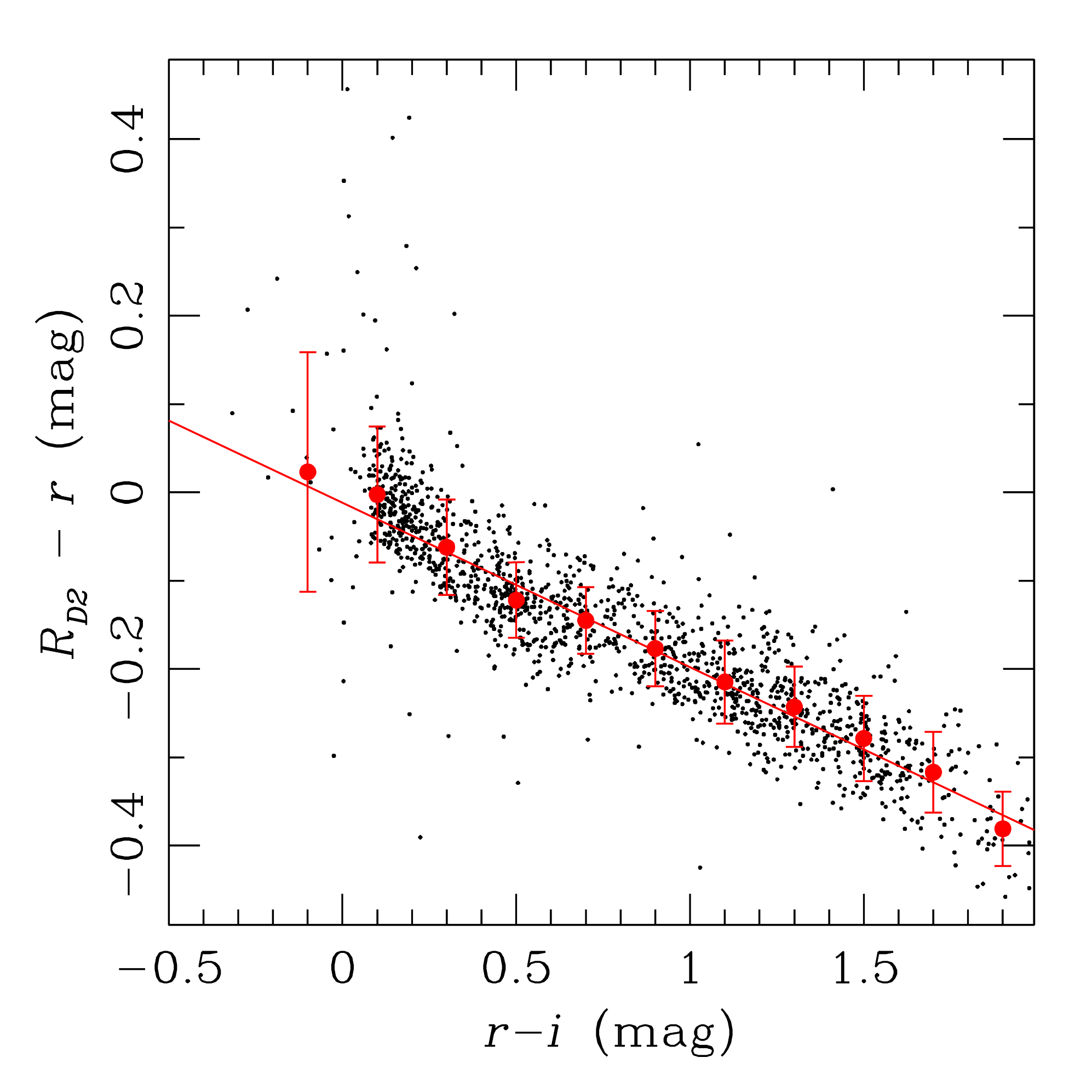}
\end{center}
\caption{$R_{\rm D2}-r$ plotted against  $r-i$ colour.  The black dots
are unresolved sources in the  D3 region of the CFHTLS (parameter {\sc
flag\_terapix}  equals   unity,  Coupon  et  al.    2009)  with  DEEP2
counterparts  in the  magnitude range  $18<R_{D2}<23$\,mag.   The mean
$R_{\rm D2}-r$ and 1\,sigma rms  within $r-i$ bins (0.2\,mag size) are
shown with  the red circles and  errorbars.  The red line  is the best
linear fit to those datapoints.  }\label{fig_cfhtls_deep2_Rr}
\end{figure}

\begin{figure}
\begin{center}
\includegraphics[height=0.9\columnwidth]{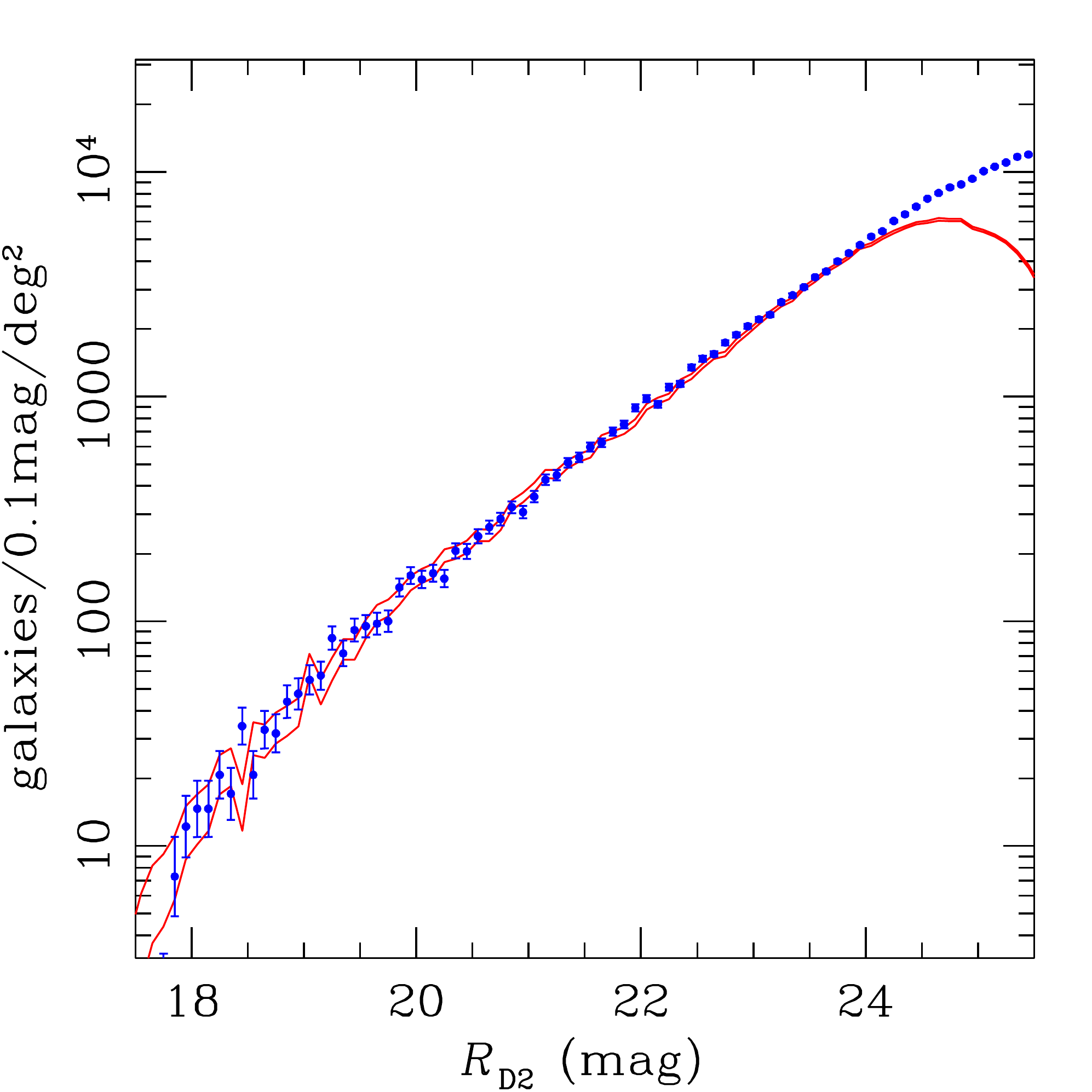}
\end{center}
\caption{Galaxy  number counts in  $R_{\rm D2}$  band. The  red curves
  correspond to the 1\,sigma rms  envelope of the galaxy counts in the
  DEEP2  photometric survey  of  the EGS.   These observations  become
  incomplete fainter than  $R\approx24$\,mag resulting in the observed
  turnover at faint magnitudes.  The blue dots are the inferred galaxy
  number counts in  the CFHTLS $r$-band converted to  the $R_{\rm D2}$
  filter using equation \ref{Rd2_to_r}}\label{fig_cfhtls_deep2_counts}
\end{figure}

\noindent The conversion from the CFHTLS $r$-band to the DEEP2 $R_{\rm
D2}$ uses optically unresolved sources (i.e.  stars).  This is because
in this case one needs  objects with accurate photometry that is least
affected by systematics  related to e.g.  the extent  of the source or
the  determination  of  aperture  corrections.   We  select  optically
unresolved sources in the CFHTLS (parameter {\sc flag\_terapix} equals
unity, \citealp  {Coupon2009}) with DEEP2  counterparts and magnitudes
in  the  range $18<R_{D2}<23$\,mag.   We  choose  not  to use  fainter
sources  to   keep  photometric  errors   small  and  also   to  avoid
contamination    by   galaxies    at   fainter    magnitudes.   Figure
\ref{fig_cfhtls_deep2_Rr}  plots $R_{\rm  D2}-r$ against  $r-i$.  Also
shown are the average and  standard deviation of $R_{\rm D2}-r$ within
$r-i$     colour    bins    of     0.2\,mag    in     size.     Figure
\ref{fig_cfhtls_deep2_Rr}  shows the best-fit  linear relation  to the
data

\begin{equation}\label{Rd2_to_r}
R_{\rm D2}-r=-0.01 - 0.19\,(r-i),
\end{equation}

\noindent The $r$-band magnitude of each galaxy in the CFHTLS D3 field
is  converted to  $R_{\rm  D2}$ and  the  corresponding galaxy  number
counts in that  filter are constructed.  These are  then compared with
the galaxy number counts estimated directly from the DEEP2 photometric
survey  of  the  EGS  field in  Figure  \ref{fig_cfhtls_deep2_counts}.
There is no evidence for systematic offsets in the two distributions.

\begin{figure*}
\begin{center}
\includegraphics[height=0.9\columnwidth]{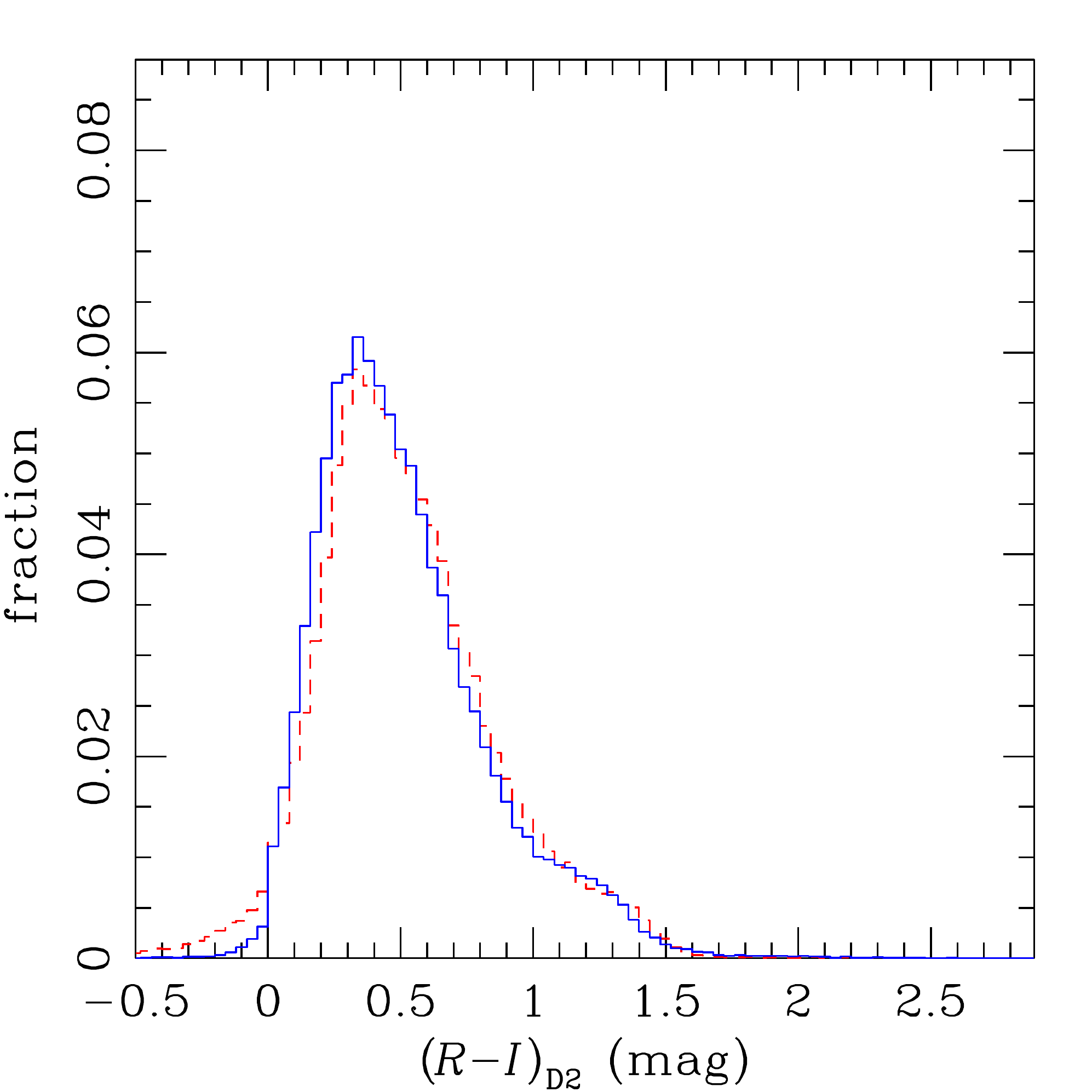}
\includegraphics[height=0.9\columnwidth]{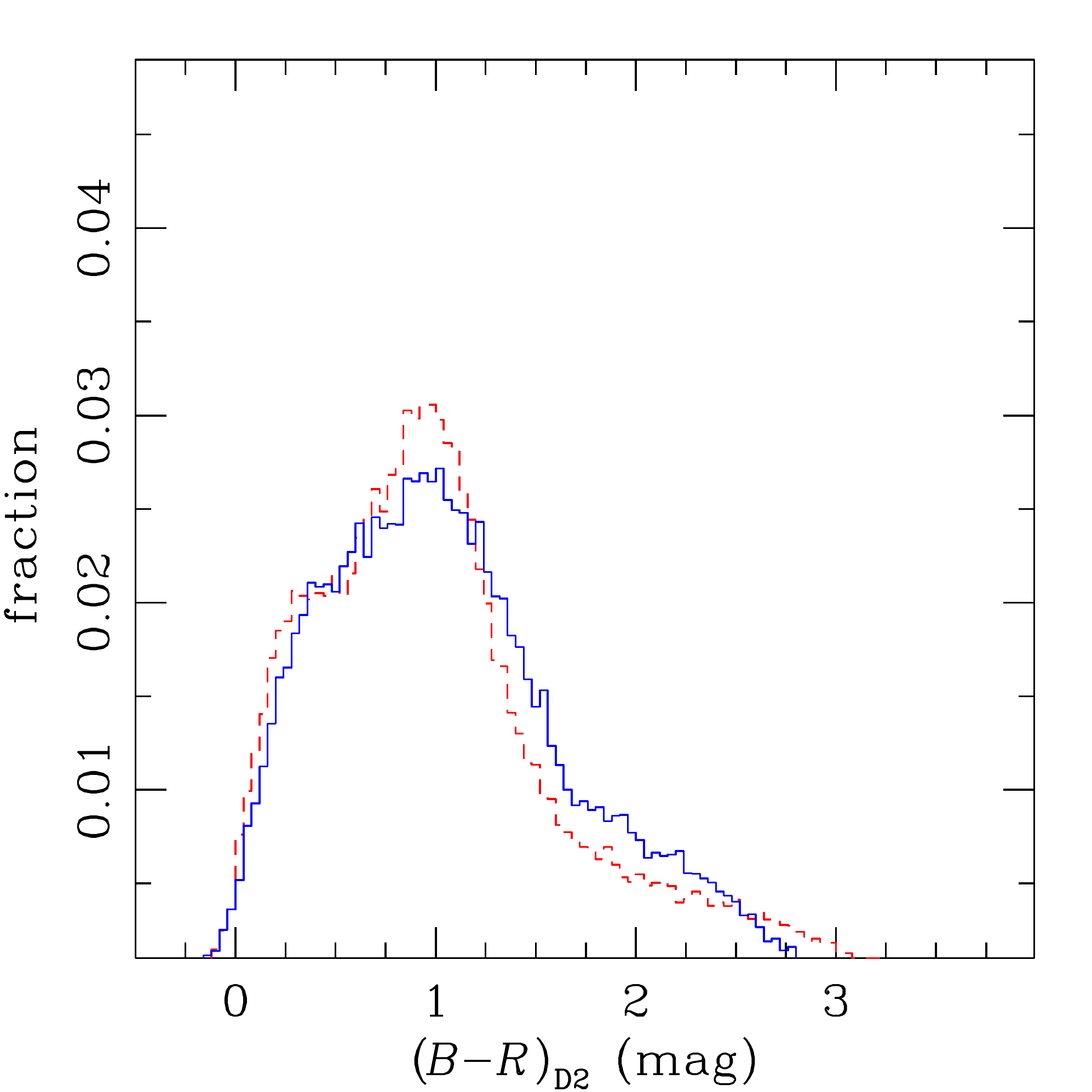}
\end{center}
\caption{$(B-R)_{\rm  D2}$ and  $(R-I)_{D2}$  colour distributions  in
  0.04\,mag  bins of galaxies  selected in  the CFHTLS-D3  (blue solid
  line) and EGS DEEP2  (red hatched histogram) surveys.  For CFHTLS-D3
  sources the  colours are estimated using  the colour transformations
  \ref{BRd2_to_gr},  \ref{RId2_to_ri}.   The  colour  distribution  is
  limited  to galaxies  in the  two surveys  with $18<R_{D2}<24$\,mag.
}\label{fig_cfhtls_deep2_colours}
\end{figure*}

We   also   investigate    the   accuracy   of   the   transformations
\ref{BRd2_to_gr}, \ref{RId2_to_ri}  by comparing the  $(B-R)_{\rm D2}$
and  $(R-I)_{\rm  D2}$ distributions  of  galaxies  inferred from  the
CFHTLS $gri$  photometry with those  estimate directly from  the DEEP2
photometric survey  in the  EGS.  In this  exercise we also  take into
account  the scatter of  datapoints around  the best-fit  relations in
Figure  \ref{fig_cfhtls_deep2}.  The  rms scatter  of  the $(B-R)_{\rm
  D2}$ versus  $g-r$ relation  is estimated to  be 0.15\,mag.  For the
$(R-I)_{\rm  D2}$ versus  $r-i$ plot  we determine  an rms  scatter of
0.1\,dex.   These numbers include  photometric errors,  differences in
the  determination  of colours  in  the two  surveys  as  well as  any
intrinsic  scatter  in  those  relations  associated  with  e.g.   the
diversity  of  the galaxy  SEDs.   For  each  CFHTLS galaxy  we  apply
equations \ref{BRd2_to_gr}, \ref{RId2_to_ri}  to infer its $(B-R)_{\rm
  D2}$ and $(R-I)_{\rm D2}$. We  then added Gaussian deviates to those
colours   with  Half  Width   Half  Maximum   of  0.15   and  0.1\,dex
respectively.   The resulting colour  distributions are  compared with
those  determined for  galaxies  in the  DEEP2  photometric survey  in
Figure  \ref{fig_cfhtls_deep2_colours}.  This  comparison is  only for
resolved  sources in  the  two surveys  with $18<R_{\rm  D2}<24$\,mag.
There  is   fair  agreement  between  the  CFHTLS   and  DEEP2  colour
distributions suggesting that at  least to the first approximation the
colour  transformations \ref{BRd2_to_gr}, \ref{RId2_to_ri}  are robust
and  can be used  to convert  the fluxes  measured through  the CFHTLS
filterset to the DEEP2 $BRI$ bands.

\section{The 2SLAQ QSO-LRG sample}

The 2dF-SDSS LRG And  QSO survey \citep{Cannon2006} is a spectroscopic
program  that used the  2dF facility  at the  AAT telescope  to follow
Luminous Red  Galaxies in the  redshift range 0.35-0.75  and UV-bright
QSOs  in  two  equatorial  strips   covering  a  total  area  of  $\rm
180\,deg^2$.  The photometric selection  of the LRG and QSO candidates
used  the  SDSS  photometry  and  the  colour  criteria  described  by
\cite{Cannon2006} and \cite{Richards2005}, respectively.

\begin{figure}
\begin{center}
\includegraphics[height=1.0\columnwidth]{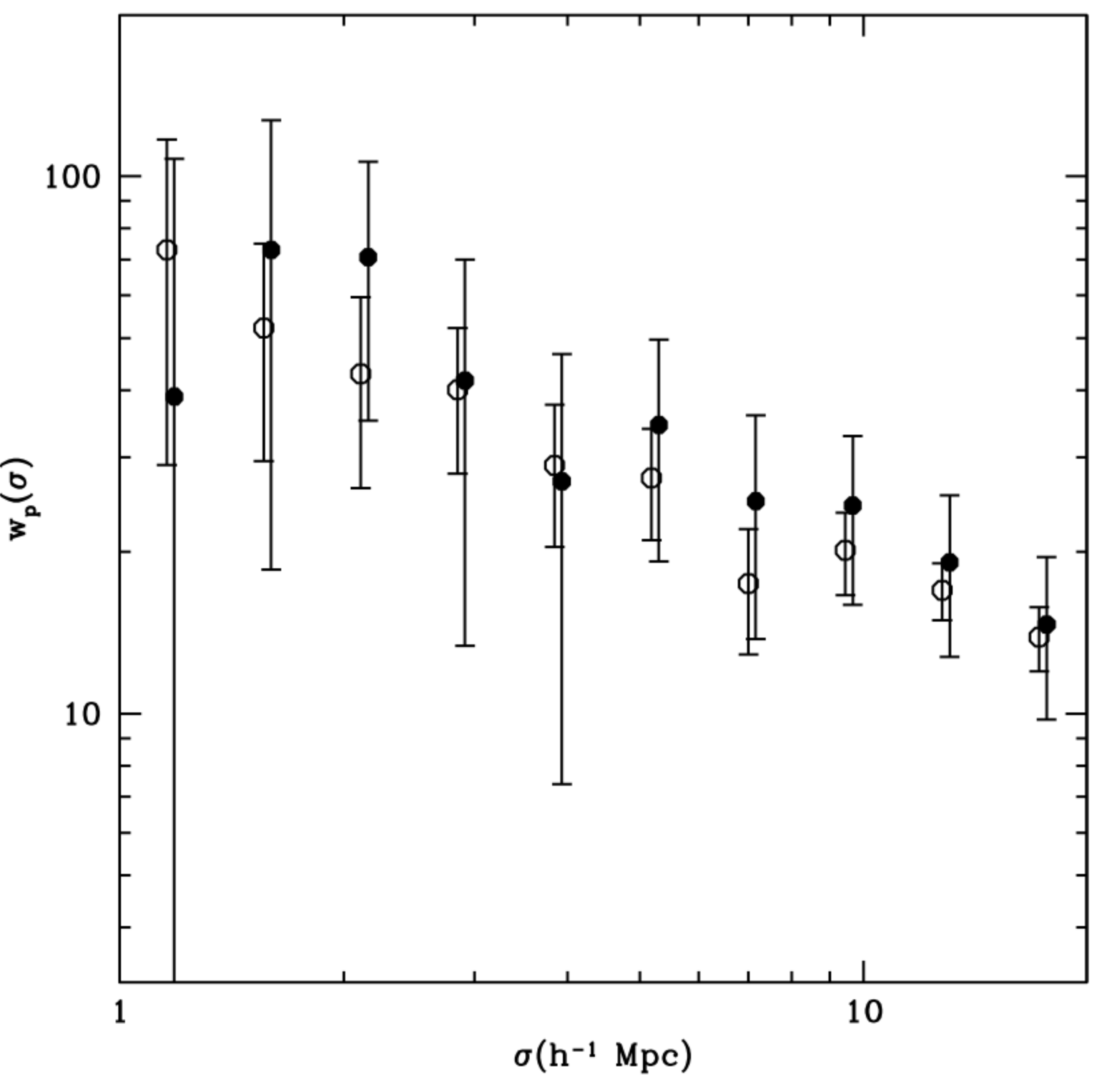}
\end{center}
\caption{QS0/LRG projected  real space cross-correlation measurements.
  Open  circles   present  the  results   when  cross-correlating  448
  spectroscopic 2SLAQ QSOs with $\sim5,500$ spectroscopic LRGs, in the
  northern strip of the 2SLAQ  survey. Filled circles plot the results
  when  the spectroscopic  LRGs have  been replaced  by  $\sim 10,000$
  photometric LRGs. Filled circles  have been offset in the horizontal
  direction by $\delta$log$\sigma$=+0.01 for clarity.}
\label{fig:qso_lrg}
\end{figure}

We  use data  from  the  northern 2SLAQ  strip,  which includes  9,923
photometrically selected LRGs and 448 spectroscopically confirmed QSOs
in the redshift interval $z=0.35-0.75$.   The LRG sample is matched to
the photometric redshift PDFs determined by \cite{Cunha2009} using the
SDSS-DR7  photometry. The QSO/LRG  cross-correlation function  is then
estimated    using    the    methodology    described    in    section
\ref{section:method}.    The    results   are   plotted    in   Figure
\ref{fig:qso_lrg}.  For comparison also  shown in  that figure  is the
QSO/LRG   cross-correlation  signal   using  spectroscopic   for  both
populations. For  this calculation we  use the 'Gold Sample'  of 2SLAQ
LRGs  \citep[total of  $5,500$  LRGs][]{Ross2007}, which  is the  most
rigorously defined 2SLAQ LRG  sample and has the highest spectroscopic
completeness. The fibre  collision effect, due to the  fact that 2SLAQ
QSO had lower priority than 2SLAQ LRGs for spectroscopic observations,
is corrected following \cite{Mountrichas2009}.

\begin{figure}
\begin{center}
\includegraphics[height=1.0\columnwidth]{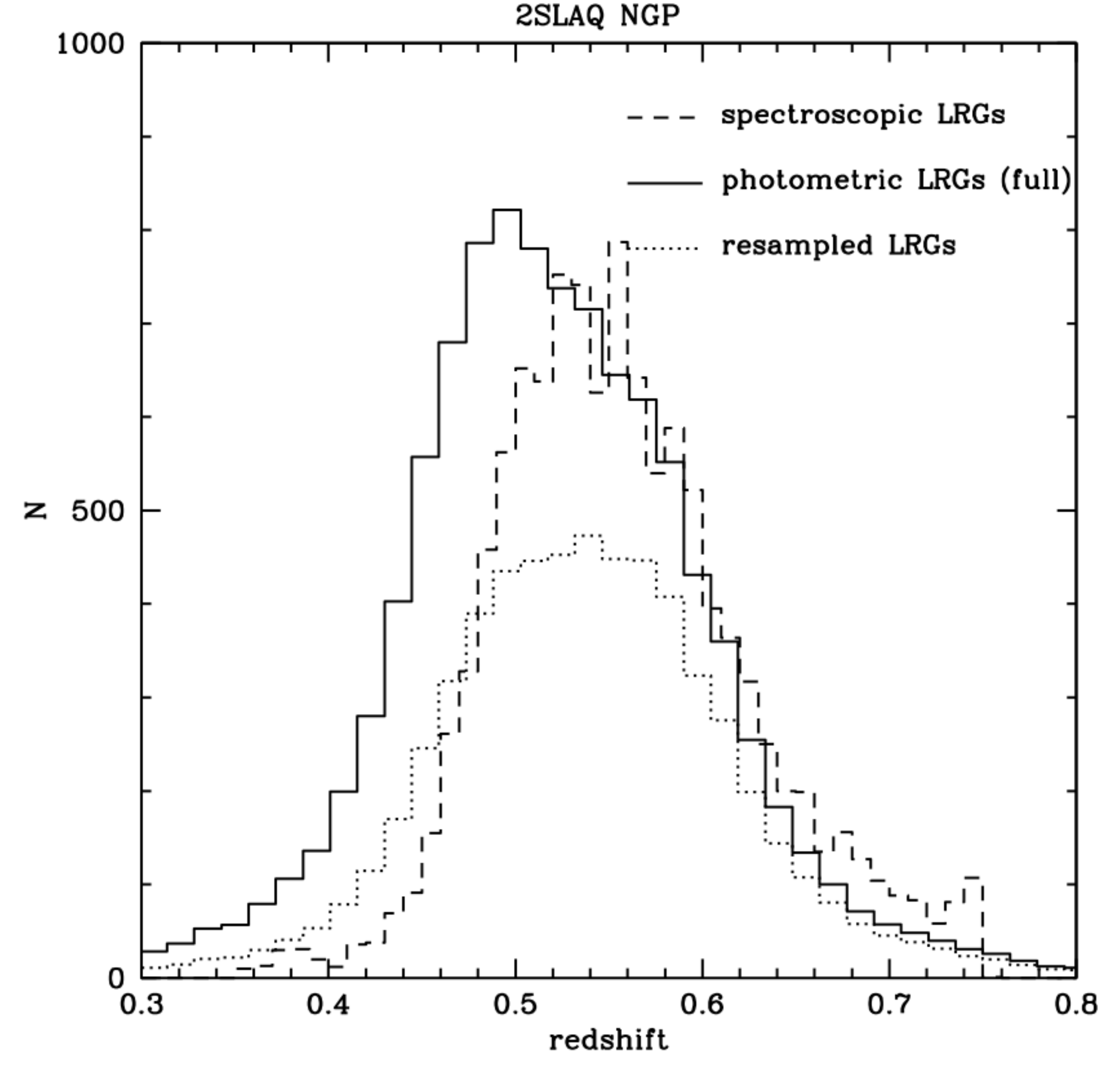}
\end{center}
\caption{The redshift distribution of the 2SLAQ LRG samples used in
  the analysis.}
\label{fig:nz_lrg}
\end{figure}

Applying  a power law  fit, on  scales $1-10\,h^{-1}$\,Mpc,  and using
equations $\ref{eqn:s8}-\ref{eqn:bs8}$, the QSO-LRG bias is estimated,
$b_{QL}=1.92^{+0.17}_{-0.15}$. The  LRG bias is  then estimated, using
(i) the  full photometric LRG  sample ($\approx 10,000$) and  (ii) the
resampling method described  in section \ref{section:method} ($\approx
6,500$  LRGs).  The  redshift  distributions of  the  two samples  are
plotted  in  Figure  $\ref{fig:nz_lrg}$.   The inferred  QSO  bias  is
$b_Q=1.72^{+0.47}_{-0.44}$          (full          sample)         and
$b_Q=1.58^{+0.50}_{-0.45}$  (resampling method).  These  estimates are
consistent within the uncertainties.



\end{document}